\documentclass[12pt,a4paper]{article}

\usepackage[utf8]{inputenc}
\usepackage[T1]{fontenc}
\usepackage{lmodern}
\usepackage{microtype}
\usepackage[title]{appendix}

\usepackage[
  a4paper,
  top=2.2cm,
  bottom=2.4cm,
  left=2.5cm,
  right=2.5cm
]{geometry}

\usepackage{amsmath,amssymb,amsthm,empheq}
\usepackage{mathrsfs}
\usepackage{graphicx}
\usepackage{float}
\usepackage[
  justification=centering,
  singlelinecheck=false
]{caption}
\usepackage[all,2cell]{xy}
\usepackage[dvipsnames]{xcolor}
\usepackage[hidelinks]{hyperref}

\DeclareMathOperator{\sech}{sech}

\newcommand{\beq}{\begin{equation}}
\newcommand{\eeq}{\end{equation}}

\newtheorem{proposicion}{Proposition}
\newtheorem{teorema}{Theorem}
\newtheorem{lema}{Lemma}

\newtheorem{remark}{Remark}

\title{Symmetries, Lax pair, and nilpotent Darboux transformation for the KdV-ghost system}

\author{%
A. Restuccia$^{1}$, J. Rosales-G\'omez$^{2}$
and A. Sotomayor$^{3}$\\[0.7em]
\small $^{1}$Departamento de F\'isica,
Universidad de Antofagasta, Chile\\
\small $^{2}$Facultad de Ciencias de la Ingenier\'ia,
Universidad Austral de Chile, Chile\\
\small $^{3}$Departamento de Matem\'aticas,
Universidad de Antofagasta, Chile\\[0.5em]
\footnotesize
\texttt{alvaro.restuccia@uantof.cl,
javier.rosales@uach.cl}\\
\footnotesize
\texttt{adrian.sotomayor@uantof.cl}
}

\date{}

\begin{document}

\raggedbottom
\maketitle

\begin{abstract}

We analyze the KdV--ghost system arising from the BRST formulation of
BF-type topological field theory, its symmetries, Lax pair, spectral
problem, zero-curvature formulation and Darboux formalism. The system
is invariant under shifts of the ghost field, whose generator does not
commute with the BRST symmetry of the KdV--ghost system. Their
generators allow the introduction of a systematic procedure for
generating new solutions from a given one, and the corresponding
expressions in the Grassmann algebra are obtained explicitly. We
introduce a Grassmann-valued Lax pair and prove an isospectral theorem
relating the spectral problem of the spatial Lax operator for the KdV
equation to that of its KdV--ghost extension. We also construct a
triangular zero-curvature representation and show that the
Grassmann-valued spatial Lax operator is related to the KdV spatial
Lax operator by a nilpotent conjugation. The natural variable for a
local Darboux formulation is the differentiated ghost $v=C_x$, for
which the ghost equation coincides with the Fr\'echet linearization of
the KdV flow. We then construct a nilpotent lift of the scalar KdV
B\"acklund--Darboux transformation. Using Riccati variables and
off-shell factorization identities for the KdV residue and its
Fr\'echet derivative, we prove that the transformation closes on the
local KdV--ghost system. Successive Darboux transformations yield a
recursive ghost extension of the bosonic Darboux chain, with a new odd
homogeneous contribution introduced at each step. For the solitonic
examples, the resulting local ghost fields are naturally identified
with tangent directions associated with the soliton phase parameters.
The construction provides a Darboux-theoretic interpretation of the
ghost sector while remaining distinct from the standard
supersymmetric KdV hierarchies.

\end{abstract}
	

	Pacs: 02.30.lk; 11.15.Ex; 02.30.Jr; 11.30.-j

\section{Introduction}
\label{sec:introduction}

The Korteweg--de Vries (KdV) equation is a paradigmatic example of an
integrable nonlinear evolution equation.
Its combination of nonlinear steepening and dispersion gives rise to
stable solitary waves and, more generally, to a rich integrable
hierarchy. Beyond its original hydrodynamical interpretation, the KdV
equation has played a central role in mathematical physics through its
Lax representation, infinite families of conserved quantities,
Hamiltonian structures, spectral theory, and connections with
algebraic geometry and infinite-dimensional symmetry algebras \cite{Gardner1967,Lax1968,Miura1968}.

A geometric approach to integrability is provided by zero-curvature
representations, in which nonlinear evolution equations arise as
compatibility conditions for auxiliary linear systems
\cite{ZakharovShabat1974,FaddeevTakhtajan1987}. This perspective is
particularly natural in gauge-theoretic formulations, where
zero-curvature conditions can be interpreted in terms of flat
connections and integrable equations may arise through suitable
reductions or gauge-fixing conditions. In this direction, Fukuyama,
Kamimura, Kre\v{s}i\'c-Juri\'c, and Meljanac studied several
integrable systems in zero-curvature form within an
$SL(2,\mathbb{R})$-invariant gauge theory and obtained a family
containing, among other systems, the KdV and Harry Dym equations,
together with gauge-equivalent modified KdV and Calogero--KdV
equations \cite{FukuyamaKamimuraKresicJuricMeljanac2007}. The
Painlev\'e properties and further integrable structures of the
corresponding FKKM hierarchy were subsequently investigated in
\cite{RestucciaRosalesGomez2026}.

A broader gauge-theoretic setting is provided by topological BF
theories, whose field equations are naturally related to flat
connections. In their BRST formulation, gauge symmetry is encoded by
a nilpotent cohomological differential, allowing one to extend the
classical field content by Grassmann-odd ghost variables
\cite{BecchiRouetStora1976}. This provides a natural framework in
which integrable dynamics, gauge geometry, and cohomological
structures can be studied simultaneously.

In \cite{RestucciaSotomayor}, different partial gauge-fixing conditions imposed on
the zero-curvature equations of a topological BF theory with structure
group $SL(2,\mathbb{R})$ led to a class of BRST-invariant extensions of
integrable equations, including the KdV, modified KdV, Calogero--KdV,
and Harry Dym equations. In particular, the resulting KdV--ghost
system couples the usual bosonic KdV field to a Grassmann-odd ghost
field. The odd component should be distinguished from the fermionic
fields appearing in supersymmetric extensions of KdV: here it
originates from the gauge and BRST structure rather than from
spacetime supersymmetry. Nevertheless, the ghost field has a
nontrivial evolution equation and participates in the integrable
structure of the extended system.

The Hamiltonian and Poisson structures of the KdV--ghost hierarchy,
together with its BRST properties and conserved quantities, were
investigated in \cite{RestucciaSotomayor2025}. These results established the
integrable character of the extended hierarchy and showed that the
BRST structure is compatible with its Hamiltonian formulation. The
purpose of the present work is to develop further the integrable
structure of the KdV--ghost system from a different direction,
focusing on its symmetries, spectral representation, tangent dynamics,
and Bäcklund--Darboux transformations.

Our first observation is the existence of a nontrivial shift symmetry
acting on the ghost field. This transformation does not commute with
the BRST symmetry. Instead, their commutator closes on a spatial
translation. This algebraic relation is reminiscent of the closure
relations encountered in supersymmetric systems, although the present
construction is not a supersymmetric extension of KdV: the odd
transformation is associated with the BRST ghost sector. This
observation motivates the construction of a successive
BRST--shift--BRST transformation scheme, which generates new
solutions from a given solution of the KdV--ghost system.

We next construct a Lax representation and a zero-curvature
formulation for the local KdV--ghost system. A central feature of the
construction is that the spatial Lax operator of the extended system
is related to the ordinary KdV spatial operator by a nilpotent gauge
transformation. This relation allows us to compare directly the
corresponding spectral problems. In particular, we establish an
isospectral correspondence between the bosonic KdV operator and its
ghost-extended counterpart. Thus, the odd extension does not introduce
an independent spectral problem, but is incorporated into the KdV
spectral structure through a nilpotent gauge transformation.

The spectral relation also provides a natural connection with the
tangent dynamics of KdV. Linearizing the KdV equation around a
solution produces the corresponding variational equation, whose
solutions describe tangent directions to the space of KdV solutions.
We show that the ghost component of the KdV--ghost system can be
interpreted within this tangent framework. For solitonic solutions,
this interpretation becomes particularly transparent: the local
ghost fields generated by the construction are naturally associated
with tangent directions corresponding to variations of the soliton
parameters, in particular their phase parameters. This gives a
geometric interpretation of the ghost sector that complements its
original BRST origin.

A further objective of this work is to establish a Darboux-theoretic
description of the ghost extension. Starting from the scalar KdV B\"acklund--Darboux transformation
\cite{MatveevSalle1991,GuHuZhou2005}, we construct a nilpotent lift
acting on the KdV--ghost system. The construction is based on Riccati
variables together with off-shell factorization identities for the KdV
residue and its Fréchet derivative. These identities allow us to prove
that the Darboux transformation closes on the local KdV--ghost system,
rather than requiring nonlocal auxiliary fields. Iterating the
transformation produces a recursive ghost extension of the ordinary
bosonic Darboux chain, with a new odd homogeneous contribution
appearing at each step. The resulting construction gives a direct
relation between the Darboux transformation of KdV and the generation
of its BRST-ghost sector.

The combination of these results establishes several complementary
links between the KdV--ghost system and the ordinary KdV hierarchy:
\begin{equation*}
\text{BRST symmetry}
\quad\longleftrightarrow\quad
\text{shift symmetry},
\end{equation*}
\begin{equation*}
\text{nilpotent gauge transformation}
\quad\longleftrightarrow\quad
\text{isospectral Lax representation},
\end{equation*}
and
\begin{equation*}
\text{Darboux transformation}
\quad\longleftrightarrow\quad
\text{recursive ghost extension}.
\end{equation*}
These relations provide an integrable and geometric characterization
of the ghost sector and distinguish the construction from the
standard supersymmetric KdV hierarchies.

The paper is organized as follows. In Section~\ref{sec:local-system}, we
review the geometric construction of the KdV--ghost system, introduce
the ghost shift symmetry, and analyze its algebra with the BRST
transformation. We then construct the successive
BRST--shift--BRST transformations and discuss their action on
solutions. In Section~3, we construct the spectral problem and the Lax
representation of the KdV--ghost system and establish its relation
with the ordinary KdV spectral problem. Section~4 is devoted to the
tangent dynamics of KdV and its relation to the ghost sector,
together with the triangular zero-curvature formulation. In
Section~5, we construct the nilpotent Bäcklund--Darboux
transformation and prove its closure on the local KdV--ghost system.
Section~6 contains explicit solitonic examples and their recursive
Darboux extensions. Finally, Section~7 discusses the relation with
integrable couplings and supersymmetric KdV and summarizes the
geometric interpretation of the ghost sector.

\section{The KdV--ghost system and the BRST and shift symmetries}
\label{sec:local-system}

In this section, we recall the KdV--ghost equations and their BRST
symmetry, and introduce a second, nontrivial symmetry given by a shift
of the ghost field. We analyze the algebra generated by the BRST and
shift transformations and show that their commutator closes on a
spatial translation. This relation provides the starting point for a
successive BRST--shift--BRST construction, which allows new solutions
of the KdV--ghost system to be generated from a given one.

We begin by recalling the KdV--ghost system and its BRST
transformations. We then define the shift symmetry and verify its
invariance properties. The resulting transformation algebra is
analyzed explicitly, and the successive BRST--shift--BRST
construction is presented at the end of the section.

\subsection{The BRST-induced KdV--ghost system}
\label{subsec:original-ghost-system}

The KdV--ghost system arising from the BRST formulation of the
two-dimensional non-Abelian BF theory is given by
\begin{equation}
\label{eq:kdv-ghost-original}
\begin{cases}
u_t=u_{xxx}+3uu_x,\\[1mm]
\widetilde C^{+}_t
=
\widetilde C^{+}_{xxx}
+3u\widetilde C^{+}_x.
\end{cases}
\end{equation}
Here $u=u(x,t)$ is an even field and
$\widetilde C^+=\widetilde C^+(x,t)$ is an odd field,
both valued in a Grassmann algebra~ \cite{RestucciaSotomayor,RestucciaSotomayor2025}.

The system \eqref{eq:kdv-ghost-original} is invariant under the BRST
transformations
\begin{equation}
\label{eq:brst-transformations}
\begin{cases}
\widehat{\delta}\widetilde C^{+}
=
\widetilde C^{+}\widetilde C^{+}_x,\\[1mm]
\widehat{\delta}u
=
\widetilde C^{+}_{xxx}
+u_x\widetilde C^{+}
+2u\widetilde C^{+}_x.
\end{cases}
\end{equation}
For notational simplicity, from now on we write
\(
C:=\widetilde C^{+}.
\)

Moreover, the system possesses an infinite sequence of conserved
quantities, including quantities depending explicitly on the ghost
field.

The formulation in terms of \(\widetilde C^{+}\) is natural from the
BRST point of view. Although the ghost equation is linear on the ghost field, the BRST transformation is not linear on it.

However, for the construction of a local
Bäcklund--Darboux transformation, where one considers the tangential dynamics of the system, it is more convenient to work with the differentiated ghost field.

\subsection{Geometric interpretation of the KdV-ghost system and its BRST invariance}
\label{subsec:evo vector fields and BRSR invariance}

Let us consider the KdV equation
\begin{equation*}
\label{eq:kdv-vector-field}
D_tu= K[u], \hspace{5mm}K[u]=D_x^3u+3uD_xu,
\end{equation*}
where \(
D_t=\dfrac{\partial}{\partial t},
\quad
D_x=\dfrac{\partial}{\partial x}.
\)

The KdV equation determines a flow map of integral curves associated
with the vector field \(K[u]\), defined on the infinite-dimensional
space of real functions \(u(x)\). We assume that \(
u\in\mathcal{S}(\mathbb{R}),
\) the Schwartz space of functions whose derivatives of all orders are
rapidly decreasing. In general, \([u]\) denotes an argument depending
on \(u\) and a finite number of \(x\)-derivatives; in the case of
\(K[u]\), the number of derivatives is three.

We have
\begin{equation*}
\label{eq:formal-kdv-vector-field}
X[u]
=
K[u]\frac{\delta}{\delta u}.
\end{equation*}

The exact expression of the vector field \(X[u]\) acting on functions
\(f[u]\) is
\begin{equation}
\label{eq:exact-kdv-vector-field}
X[u]f[u]
=
\sum_{\ell=0}^{m}
D_x^\ell K[u]\,
\frac{\partial}{\partial u_\ell}
f(u_0,u_1,\ldots,u_m),
\end{equation}
where one replaces \(
D_x^\ell u\longrightarrow u_\ell
\) in the argument of \(f\), perform the derivatives, and then replaces
back \(
u_\ell\longrightarrow D_x^\ell u.
\)
For example, considering \(u_\ell\), \(\ell=1,\ldots,m\), as
independent variables,
\[
X[u]u=K[u],
\qquad
X[u]D_xu=D_xK[u].
\]
We then have
\[
[D_x,X]u
=
D_xXu-XD_xu
=
D_xK[u]-D_xK[u]
=
0.
\]
The same result follows for any \(f[u]\):
\begin{equation}
\label{eq:Dx-X-commutator}
[D_x,X]f[u]=0.
\end{equation}

That is, \(D_x\) and \(X\) commute.

The extension of the KdV equation, the KdV--ghost system, can be
interpreted as a flow map of integral curves on the space of functions
with values in a Grassmann algebra, a superspace, where \(u\) belongs
to the even part of the algebra and \(C\) to the odd part. Hence \(C\)
is nilpotent, while \(u\) decomposes into the body and the even
nilpotent part. Hence the KdV--ghost system is now extended to
Grassmann-valued fields.

The vector field is now formulated as
\begin{equation*}
\label{eq:extended-evolutionary-vector-field}
\widetilde{X}[u,C]
=
K[u]\frac{\delta}{\delta u}
+
\mathcal{Y}[u,C]\frac{\delta}{\delta C},
\end{equation*}
where \(
\mathcal{Y}[u,C]
=
D_x^3C+3uD_xC.
\)

The flow of integral curves follows from the KdV--ghost system,
\begin{equation*}
\label{eq:extended-flow}
D_tu=\widetilde{X}[u,C]u,
\qquad
D_tC=\widetilde{X}[u,C]C.
\end{equation*}

In \cite{RestucciaSotomayor2025} we introduced a BRST vector field
\begin{equation}
\label{eq:brst-vector-field}
\Phi[u,C]
\equiv
\left(
D_x^3C+2uD_xC+CD_xu
\right)
\frac{\delta}{\delta u}
+
CC_x\frac{\delta}{\delta C},
\end{equation}
where
\(
P[u]C
\equiv
D_x^3C+2uD_xC+CD_xu
\)
and \(CC_x\) are the BRST transforms of \(u\) and \(C\), respectively.
The explicit expression of \eqref{eq:brst-vector-field} is given by a
formal extension of \eqref{eq:exact-kdv-vector-field}, where derivatives
with respect to \(C\) and its \(x\)-derivatives are also included in the
summation on the right-hand side of \eqref{eq:exact-kdv-vector-field}.

\begin{lema}
Let \(u\) and \(C\) belong to a Grassmann algebra \(\mathcal{G}\)
generated by a finite number of elements
\(\theta_1,\ldots,\theta_s\), \(s>1\), with \(u\) belonging to the even
part of \(\mathcal{G}\) and \(C\) to the odd part of \(\mathcal{G}\).
Assume that \(u\) and \(C\) belong to
\(\mathcal{S}(\mathbb{R})\), in the sense that every coefficient with
respect to the basis generated by
\(\theta_1,\ldots,\theta_s\) belongs to
\(\mathcal{S}(\mathbb{R})\). Then
\begin{equation}
\label{eq:X-Phi-commutator}
[\widetilde{X},\Phi]=0.
\end{equation}
\end{lema}

\begin{proof}
We first notice that, besides \(
[D_x,\widetilde{X}]=0,
\) one also has
\begin{equation}
\label{eq:Dx-Phi-commutator}
[D_x,\Phi]=0.
\end{equation}
In fact, the same argument which yields
\eqref{eq:Dx-X-commutator} also applies to
\eqref{eq:Dx-Phi-commutator}.

In order to prove the lemma, it is enough to verify
\begin{align}
[\widetilde{X},\Phi]u&=0,
\label{eq:X-Phi-u}\\
[\widetilde{X},\Phi]C&=0.
\label{eq:X-Phi-C}
\end{align}

Let us check \eqref{eq:X-Phi-u}. We have
\[
\widetilde{X}u=K[u],
\qquad
\Phi u=P[u]C.
\]
Then, using \eqref{eq:Dx-X-commutator}, we get
\begin{align}
\widetilde{X}\Phi u
&=
D_x^3\widetilde{X}C
+
u_x\widetilde{X}C
+
2D_xC\cdot\widetilde{X}u
+
2uD_x\widetilde{X}C
\nonumber\\
&=
D_x^3\mathcal{Y}[u,C]
+
\bigl(D_x^4u+D_x(3uu_x)\bigr)C
+
u_x\bigl(C_{xxx}+3uC_x\bigr)
\nonumber\\
&\quad
+
2\bigl(u_{xxx}+3uu_x\bigr)C_x
+
2u\bigl(C_{xxxx}+3u_xC_x+3uC_{xx}\bigr).
\label{eq:X-Phi-u-first}
\end{align}
On the other hand,
\begin{align}
\Phi\widetilde{X}u
&=
D_x^3\mathcal{Y}
+
D_x^4u\cdot C
+
u_xD_x^3C
+
3D_x^3u\cdot D_xC
+
3D_x^2u\cdot D_x^2C
\nonumber\\
&\quad
-
\Bigl(
D_x^3u\cdot D_xC
+
3D_x^2u\cdot D_x^2C
+
3D_xu\cdot D_x^3C
+
uD_x^4C
\Bigr)
\nonumber\\
&\quad
+
3\bigl(D_x^3C+CD_xu+2uD_xC\bigr)D_xu
\nonumber\\
&\quad
+
3uD_x\bigl(D_x^3C+CD_xu+2uD_xC\bigr).
\label{eq:Phi-X-u}
\end{align}

Finally, comparing each coefficient of the \(x\)-derivatives of \(C\),
we obtain (8). A similar calculation yields
\eqref{eq:X-Phi-C}.

The lemma implies that the BRST transformation maps solutions into
solutions of the KdV--ghost system.
\end{proof}

In contrast to the vector field $\widetilde{X}$ which is linear in the ghost field $C$, the vector field $\Phi$ is non-linear in $C$. This property ensures that $\Phi$ is not contained in the tangent dynamics of the KdV geometry.

If $(u,C)$ is a solution of the KdV-ghost system, which may be rewritten as
\begin{equation}\begin{array}{ll}D_tu-\widetilde{X}u=0,\\
D_tC-\widetilde{X}C=0.\end{array}\label{1}
\end{equation}

In fact, if (\ref{1}) are valid we get

\begin{equation*}\begin{array}{ll}\Phi(D_tu-\widetilde{X}u)=0,\\
\Phi(D_tC-\widetilde{X}C)=0,\end{array}\label{2}
\end{equation*}
where we have simplified the notation and used $\Phi$ instead of $\Phi[u,C]$ and $\widetilde{X}$ instead of $\widetilde{X}[u,C].$

Therefore,

\begin{equation}\begin{array}{ll}\Phi(D_tu-\widetilde{X}u)=D_t\Phi u-\Phi
\widetilde{X}u=D_t\Phi u-\widetilde{X}\Phi u=0,\\
\Phi(D_tC-\widetilde{X}C)=D_t\Phi C-\Phi
\widetilde{X}C=D_t\Phi C-\widetilde{X}\Phi C=0.\end{array}\label{3}
\end{equation}

Equation (11) expresses the infinitesimal symmetry condition generated by $\Phi$.
It does not imply that $(\Phi u,\Phi C)$ itself is a solution of the nonlinear
KdV--ghost system. Rather, if $(u,C)$ is a solution, then the BRST-transformed fields
\[
u' = u + \epsilon\,\Phi u,
\qquad
C' = C + \epsilon\,\Phi C,
\]
where $\epsilon$ is an odd nilpotent parameter, are again a solution. Since
$\epsilon^2=0$, the transformation is exact. This is the form of the BRST
transformation used below in the construction of new solutions.

In deriving (11), we have used the commutation relation (6), together with the evolution equations (10).

The other relevant remark is that the KdV-ghost system is also invariant under the shift transformation $T$ given by $C\rightarrow C+\mathbb{C}$, where $\mathbb{C}$ is a constant odd quantity and $u\rightarrow u$.

The generators of the two symmetries do not commute; therefore, we can combine them
to generate new solutions from known ones.

The main point, in the generation of new solutions from known ones,  is that the generator $\Phi[u,C]$ of the BRST transformation and the generator  $S_{\mathbb{C}}[C]$ of the $T$-transformation do not commute. It is interesting that their commutator generates a spatial translation. In fact,
\[ S_{\mathbb{C}}u=0,\hspace{5mm} S_{\mathbb{C}}C=\mathbb{C},\]
for $\mathbb{C}$ an arbitrary constant odd parameter, and hence we can represent ${\mathbb{C}}$ as
\[{S_{\mathbb{C}}}=\mathbb{C}\frac{\delta}{\delta C}.\]

We then have
\begin{eqnarray*}&& [S_{\mathbb{C}},\epsilon \Phi]u=
S_{\mathbb{C}}\epsilon \Phi u-\epsilon \Phi_{\mathbb{C}}u=\epsilon S_{\mathbb{C}}P[u]C=
\epsilon P[u]\mathbb{C}=\epsilon \mathbb{C}u_x\\
&& [S_{\mathbb{C}},\epsilon \Phi]C=
S_{\mathbb{C}}\epsilon C C_x-\epsilon \Phi S_{\mathbb{C}}C=
\epsilon \mathbb{C}C_x-\epsilon \Phi\mathbb{C}=\epsilon \mathbb{C}C_x, \end{eqnarray*}
therefore \begin{equation}[S_{\mathbb{C}}, \epsilon \Phi]=\epsilon \mathbb{C}T_x, \label{4}\end{equation}
where $T_x=u_x\frac{\delta}{\delta u}+C_x\frac{\delta}{\delta C}$ is the generator of spatial translations.

We also have

\begin{equation}[S_{\mathbb{C}},T_x]=0,\hspace{5mm}[\Phi,T_x]=0.\label{5}\end{equation}
(\ref{4}) is reminiscent of the supersymmetric algebra where the anticommutator of two supersymmetric generators generates a translation. In (\ref{4}) we have already incorporated the odd parameters $\epsilon$ and $\mathbb{C}$.

The relevance of the algebra (\ref{4}) and (\ref{5}) is that we can use a sequence of BRST and $T$-transformations in order to generate new solutions from known ones
\[\cdots\rightarrow (u,C)\rightarrow_{BRST}(u^\prime,C^\prime)\rightarrow^T
(u^{\prime\prime},C^{\prime\prime})\rightarrow_{BRST}\dots\]
Since the BRST transformation is nilpotent, if $S_{\mathbb{C}}$ and $Q$ had commuted, the sequence would have stopped at the first step.

We will discuss the generation of solutions in next section.

\subsection{New solutions from the BRST-$T$-BRST composition}
Let $(u_0,C_0)$ be a solution of the KdV-ghost system.

Then
\[u_1=u_0+\epsilon \hat{\delta}u_0, \hspace{5mm}C_1=C_0+\epsilon \hat{\delta}C_0\] is also a solution.

We now perform a ghost shift transformation,
\(
u_2=u_1,\quad C_2=C_1+\gamma_1,
\)
where $\gamma_1$ is an $x$- and $t$-independent odd constant of the Grassmann algebra.
Finally, we perform a new BRST transformation,
\[
u_3=u_2+\widetilde{\epsilon}\,\hat{\delta}u_2,
\qquad
C_3=C_2+\widetilde{\epsilon}\,\hat{\delta}C_2.
\]
The sequence may continue and new solutions will arise.

Here, we will analyze $(u_3,C_3)$, obtain their explicit expression and discuss its Grassmann structure.

We have,
\begin{eqnarray*}&& u_3=u_0+\epsilon \hat{\delta}u_0+\tilde{\epsilon}
\hat{\delta}u_2, \hspace{3mm}C_3=C_0+\epsilon\hat{\delta}C_0+\gamma_1+\tilde{\epsilon}\hat{\delta}C_2,\\&&\hat{\delta}u_2=P[u_2]C_2=P[u_0+\epsilon\hat{\delta}u_0
](C_0+\epsilon\hat{\delta}C_0+\gamma_1), \\&& \hat{\delta}C_2
=C_2C_{2x}=(C_0+\epsilon\hat{\delta}C_0+\gamma_1){(C_0+\epsilon\hat{\delta}C_0)}_x,   \end{eqnarray*}
where $\hat{\delta}u_0$ and $\hat{\delta}C_0$ are given by the
BRST transformations~(5).

After some calculations we obtain
\begin{eqnarray*}&&u_3=u_0+(\epsilon+\tilde{\epsilon}) \hat{\delta}u_0+\tilde{\epsilon}\gamma_1u_{0x}-\tilde{\epsilon}\epsilon\gamma_1
{(\hat{\delta}u_0)}_x,\\&& C_3=C_0+\gamma_1+(\epsilon+\tilde{\epsilon}) \hat{\delta}C_0+\tilde{\epsilon}\gamma_1C_{0x}-\tilde{\epsilon}\epsilon
\gamma_1{(\hat{\delta}C_0)}_x. \end{eqnarray*}

By the BRST and shift invariance established above, $(u_3,C_3)$ is
a solution of the KdV--ghost equations. We have also checked this explicitly.

If $\gamma_1=0$ then the solutions reduce to the expressions
\begin{eqnarray*}&& u_3|_{\gamma_1=0}=u_0+(\epsilon+\tilde{\epsilon})\hat{\delta}u_0,\\&&
C_3|_{\gamma_1=0}=C_0+(\epsilon+\tilde{\epsilon})\hat{\delta}C_0, \end{eqnarray*}
the same ones that are obtained after performing only one BRST transformation, as expected.

If $\gamma_1\neq 0$, the composition generally yields new solutions. In order to satisfy the KdV-ghost equations all terms in the above expression have to be included. Due to the non-linearity of the equations the terms with different parameters, for example $\epsilon\hat{\delta}u_0$ and
$\tilde{\epsilon}\gamma_1u_{0x}$ contribute to the terms with parameter
$\tilde{\epsilon}\epsilon\gamma_1$ in the equation for $u$.

The same occurs with the terms $\epsilon\hat{\delta}C_0$ and
$\tilde{\epsilon}\gamma_1C_{0x}$ which combine with $\epsilon\hat{\delta}u_0$ and
$\tilde{\epsilon}\gamma_1u_{0x}$ yielding a contribution to the terms with parameter $\tilde{\epsilon}\epsilon\gamma_1$ in the ghost equation.

In order to obtain non-trivial results for $u_3$ and $C_3$ one must consider a Grassmann algebra with at least five odd generators $\theta_1,\theta_2,\ldots,\theta_5.$

In fact, one may consider the field expanded in the base generated by $\theta_1,\theta_2:1,\theta_1,\theta_2,\theta_1\theta_2,$ and the odd parameters expressed in terms of the other three generators $\theta_3,\theta_4$ and $\theta_5.$

We need at least two generators $\theta_1$ and $\theta_2$ in order to have a non-trivial $\tilde{\delta}C=CC_x$.

\section{Lax representation
and nilpotent gauge interpretation}
\label{sec:zcr-gauge}

We begin by constructing a differential Lax representation that incorporates the ghost sector directly into the standard KdV framework. The Grassmann-valued extension is introduced through an odd nilpotent parameter, so that the bosonic KdV pair is recovered in the vanishing-parameter limit, while the nilpotent contribution encodes the ghost dynamics.

\subsection{The Grassmann-valued differential Lax pair}
\label{subsec:differential-lax-pair}

Let \(\varepsilon\) be an odd nilpotent parameter satisfying
\(
\varepsilon^2=0.
\)
Since both \(\varepsilon\) and \(C\) are odd, the product
\(\varepsilon C\) is even and nilpotent.

The KdV--ghost system admits a Grassmann-valued differential Lax
representation. The operators below are obtained from a general
differential-operator ansatz, as derived in Appendix A. The spatial
operator is
\begin{equation}
\label{eq:grassmann-spatial-lax}
\mathscr{L}_{\varepsilon}
=
D_x^2
+\varepsilon C_xD_x
+\frac{1}{2}
\left(
u+\varepsilon C_{xx}
\right)I,
\end{equation}
whereas the temporal operator is
\begin{equation}
\label{eq:grassmann-temporal-lax}
\begin{aligned}
\mathscr{M}_{\varepsilon}
={}&
4D_x^3
+6\varepsilon C_xD_x^2
+\left(
3u+6\varepsilon C_{xx}
\right)D_x +
\frac{3}{2}
\left(
u_x+\varepsilon C_{xxx}
\right)I.
\end{aligned}
\end{equation}

The corresponding Lax equation is
\begin{equation}
\label{eq:extended-lax-equation}
\partial_t\mathscr{L}_\varepsilon=[\mathscr{M}_\varepsilon,\mathscr{L}_\varepsilon].
\end{equation}
At order \(\varepsilon^0\), this equation yields the KdV equation,
whereas its component linear in \(\varepsilon\) determines the ghost
equation up to an \(x\)-independent odd function.

Equivalently, after
introducing the local variable \(v=C_x\), the Lax equation reproduces
the KdV--ghost system exactly.

\subsection{Nilpotent gauge conjugation}
\label{subsec:nilpotent-gauge}

We now compare the extended pair
$(\mathscr{L}_{\varepsilon},\mathscr{M}_{\varepsilon})$
with $(\mathscr{L}_0,\mathscr{M}_0)$ through the multiplication
operator defined below.

The operators $\mathscr{L}_{\varepsilon}$ and
$\mathscr{M}_{\varepsilon}$ defined in (14) and (15), respectively,
reduce, at $\varepsilon=0$, to the bosonic KdV Lax operators.
We therefore set
\begin{equation}
\label{eq:bosonic-lax-pair}
\mathscr{L}_0
:=
\left.\mathscr{L}_{\varepsilon}\right|_{\varepsilon=0}
=
D_x^2+\frac{1}{2}uI,
\qquad
\mathscr{M}_0
:=
\left.\mathscr{M}_{\varepsilon}\right|_{\varepsilon=0}
=
4D_x^3+3uD_x+\frac{3}{2}u_xI.
\end{equation}

Define the multiplication operator
\begin{equation}\label{GG-1}
G
:=
\exp\left(\frac{\varepsilon}{2}C\right).
\end{equation}
Since \(\varepsilon\) and \(C\) are odd, \(\varepsilon C\) is even,
and since \((\varepsilon C)^2=0\),
\begin{equation*}
G
=
I+\frac{\varepsilon}{2}C,
\qquad
G^{-1}
=
I-\frac{\varepsilon}{2}C.
\end{equation*}

\begin{proposicion}[Nilpotent gauge conjugation]
The extended spatial operator satisfies
\begin{equation}\label{PL1}
\mathscr{L}_{\varepsilon}
=
G^{-1}\mathscr{L}_0G.
\end{equation}
Moreover, the temporal operators satisfy
\begin{equation}\label{LP2}
G^{-1}\mathscr{M}_0G-G^{-1}G_t
=
\mathscr{M}_{\varepsilon}
-
\frac{\varepsilon}{2}\Theta[u,C]\,I,
\end{equation}
where
\[
\Theta[u,C]:=C_t-C_{xxx}-3uC_x.
\]
\end{proposicion}

\begin{proof}
From the explicit form of \(G\),
\[
G^{-1}D_xG
=
D_x+\frac{\varepsilon}{2}C_x.
\]
Therefore,
\[
G^{-1}D_x^2G
=
D_x^2+\varepsilon C_xD_x
+\frac{\varepsilon}{2}C_{xx},
\]
and hence
\[
G^{-1}\mathscr{L}_0G
=
D_x^2+\varepsilon C_xD_x
+\frac{\varepsilon}{2}C_{xx}
+\frac12 uI
=
\mathscr{L}_{\varepsilon}
\]

Similarly,
\[
G^{-1}D_x^3G
=
D_x^3
+\frac{3\varepsilon}{2}C_xD_x^2
+\frac{3\varepsilon}{2}C_{xx}D_x
+\frac{\varepsilon}{2}C_{xxx}.
\]

Using the explicit form of $\mathcal{M}_{\varepsilon}$ given in (15), the
previous identity yields
\[
G^{-1}\mathcal{M}_0G
=
\mathcal{M}_{\varepsilon}
+\frac{\varepsilon}{2}
\left(C_{xxx}+3uC_x\right)I.
\]
Equivalently,
\[
\mathcal{M}_{\varepsilon}
=
G^{-1}\mathcal{M}_0G
-\frac{\varepsilon}{2}
\left(C_{xxx}+3uC_x\right)I.
\]

Using also
\[
G^{-1}G_t=\frac{\varepsilon}{2}C_t,
\]
direct substitution gives
\[
G^{-1}\mathscr{M}_0G-G^{-1}G_t
=
\mathscr{M}_{\varepsilon}
-
\frac{\varepsilon}{2}\Theta[u,C]\,I.
\]

On solutions of the ghost equation, $\Theta[u,C]=0$, and the usual temporal
gauge relation is recovered. Under this assumption, the extended Lax equation
(16) reduces, through the gauge transformation (19)--(20), to the standard
KdV Lax equation.
\end{proof}

\subsection{Functional setting and isospectrality}
\label{subsec:isospectrality}

We consider a Grassmann algebra \(\Lambda\) generated by
\(\theta_1,\ldots,\theta_N\), with its standard decomposition into
even and odd sectors,
\[
\Lambda=\Lambda_0\oplus\Lambda_1.
\]
Every even element \(\lambda\in\Lambda_0\) admits a unique
decomposition
\[
\lambda=\lambda_B+\lambda_S,
\]
where \(\lambda_B\) is its body and \(\lambda_S\) is its nilpotent
soul.

We assume that $u$ and $C$ are even and odd, respectively,
Grassmann-valued functions whose coefficients in the expansion with
respect to $\theta_1,\ldots,\theta_N$ belong to the real Schwartz
space.

The domain of the operator $\mathscr{L}_\varepsilon$ is
$H^2(\mathbb{R})\otimes\Lambda$, that is, the coefficients in the
$\theta$-expansion belong to the Sobolev space $H^2(\mathbb{R})$.
Likewise, the domain of $\mathscr{L}_0$ is
$H^2(\mathbb{R})\otimes\Lambda$, where
$\mathscr{L}_\varepsilon$ is related to $\mathscr{L}_0$ by
Proposition~1, with $G$ defined in~(18).

\begin{teorema}[]\label{teo1} 
Under the assumptions above, the resolvent sets of $\mathscr{L}_{\varepsilon}$ and $\mathscr{L}_{0}$ satisfy
\[
\rho(\mathscr{L}_{\varepsilon})
=
\rho(\mathscr{L}_0).
\]
Consequently, their spectra satisfy
\[
\sigma(\mathscr{L}_{\varepsilon})
=
\sigma(\mathscr{L}_0).
\]

\end{teorema}

\begin{remark}
In fact, the spectra of $\mathscr{L}_\varepsilon$ and $\mathscr{L}_0$ are defined by
\[
\sigma(\mathscr{L}_\varepsilon)
=
\Lambda_0\setminus\rho(\mathscr{L}_\varepsilon),
\]
and
\[
\sigma(\mathscr{L}_0)
=
\Lambda_0\setminus\rho(\mathscr{L}_0),
\]
respectively. Therefore,
\[
\sigma(\mathscr{L}_\varepsilon)
=
\sigma(\mathscr{L}_0).
\]
That is, the spectra of $\mathscr{L}_\varepsilon$ and
$\mathscr{L}_0$, both with domain
$H^2(\mathbb{R})\otimes\Lambda$, coincide.
Notice that the elements of the spectrum belong to $\Lambda_0$.

The resolvent set of $\mathscr{L}_\varepsilon$ is the set of
$\lambda\in\Lambda_0$ such that
\[
\mathscr{L}_\varepsilon-\lambda I:
H^2(\mathbb{R})\otimes\Lambda
\longrightarrow
L^2(\mathbb{R})\otimes\Lambda
\]
is bijective and its inverse is bounded.
\end{remark}

\begin{proof}
We have
\[
\mathscr{L}_{\varepsilon}-\lambda I
=
G^{-1}
(\mathscr{L}_0-\lambda I)G,
\]
hence
\[
(\mathscr{L}_{\varepsilon}-\lambda I)^{-1}
=
G^{-1}
(\mathscr{L}_0-\lambda I)^{-1}G.
\]

Consequently, the inverse of one operator exists if and only if the inverse
of the other operator exists. Moreover, one operator is bounded if and only
if the other one is also bounded. This is so, under the assumption that the
Grassmann algebra valued fields $u$ and $C$ belong to the Schwartz space.
In fact, G is bounded and the product of $G$ with elements of the Hilbert space $L^2$ or $H^2$
also belongs to $L^2$ or $H^2$ respectively. We then have
\[
\rho(\mathscr{L}_{\varepsilon})
=
\rho(\mathscr{L}_0),
\]
which implies
\[
\sigma(\mathscr{L}_{\varepsilon})
=
\sigma(\mathscr{L}_0).
\]
\end{proof}

A consequence of Theorem~ \ref{teo1} is that the body of the
spectrum of $\mathscr{L}_{\varepsilon}$, $\sigma_B(\mathscr{L}_{\varepsilon})$, is equal to the body of $\mathscr{L}_0$,
\[
\sigma_B(\mathscr{L}_{\varepsilon}) = \sigma_B(\mathscr{L}_0)
\]

This result can be improved.

\begin{lema}[]\label{lem2}
The following holds:
\[
\sigma_B(\mathscr{L}_{\varepsilon}) = \sigma_B(\mathscr{L}_0) = \sigma(\mathscr{L}_{0B})
\]

where
\(
\mathscr{L}_{0B}
=
D_x^2+\frac{1}{2}u_BI
\)
is the body operator on \(L^2(\mathbb{R})\), with domain
\(H^2(\mathbb{R})\).

\end{lema}

\begin{proof}[Proof of Lemma \ref{lem2}]

Let $\lambda\in\Lambda_0$ and decompose $\lambda$ and $u$ into their body
and soul parts as
\[
\lambda=\lambda_B+\lambda_S,\qquad
u=u_B+u_S.
\]
If $\lambda_B\in\rho(\mathcal L_{0B})$, then
\[
\mathscr{L}_0 - \lambda I = \Big(\mathscr{L}_{0B} - \lambda_B I \Big)\Bigg[ I - \Big(\mathscr{L}_{0B} - \lambda_B I \Big)^{-1} \left(\lambda_S - \dfrac{1}{2}u_S \right)  \Bigg]
\]
where $\ (\mathscr{L}_{0B} - \lambda_B I)^{-1}\left(\lambda_S - \dfrac{1}{2}u_S \right)$ is nilpotent. Thus $I - (\mathscr{L}_{0B} - \lambda_B I)^{-1} \left(\lambda_S - \dfrac{1}{2}u_S \right)$
is invertible. Consequently, if $\lambda_B\in\rho(\mathcal{L}_{0B})$, then, for every
nilpotent $\lambda_S$, the element $\lambda=\lambda_B+\lambda_S$ belongs to
$\rho(\mathcal{L}_0)$. Conversely, if $\lambda\in\rho(\mathcal{L}_0)$, then
its body $\lambda_B$ belongs to $\rho(\mathcal{L}_{0B})$. Therefore,
\[
\rho_B(\mathcal{L}_0)=\rho(\mathcal{L}_{0B}),
\]
and hence
\[
\sigma_B(\mathcal{L}_0)=\sigma(\mathcal{L}_{0B}).
\]

\end{proof}

\begin{remark}
Let $\hat{\lambda}$ be an isolated eigenvalue of $\mathscr{L}_{0B}$ and $P$ the spectral projection from $H^2 (\mathbb{R}) \otimes \Lambda$ to the space of eigenfunctions with eigenvalue $\hat{\lambda}$.

Let $\lambda \in \Lambda_0$ be an eigenvalue of $\mathscr{L}_0$ with body $\hat{\lambda}$:
\[
\mathscr{L}_0 \psi = \lambda \psi = (\hat{\lambda} + \lambda_S)\psi ,
\]
and assume $u_S=0$. Then, if the body of $P\psi \neq 0$, $\lambda_S = 0$.

In fact,
\[
P \mathscr{L}_0 \psi = \mathscr{L}_0 P \psi = \hat{\lambda} P \psi,
\]
then
\[
(\hat{\lambda} + \lambda_S) P \psi = \mathscr{L}_0 P \psi = \hat{\lambda} P \psi
\]

which implies $\lambda_S P \psi = 0$. Consequently,
\[
\lambda_S = 0 .
\]

Consequently, if $u_S=0$ and the body of $P\psi \neq 0$, where $\psi$ is an eigenfunction
of $\mathscr{L}_0$, then the associated eigenvalue has no soul and
it is the real eigenvalue of $\mathscr{L}_{0B}$.

\end{remark}

\section{The tangential dynamics and the zero-curvature formulation}
\label{subsec:local-variable}

We introduce the local ghost variable
\begin{equation*}
\label{eq:def-v}
v:=\partial_xC.
\end{equation*}
Taking the \(x\)-derivative of the second equation in
\eqref{eq:kdv-ghost-original}, we obtain
\begin{equation}
\label{eq:local-kdv-ghost}
\begin{cases}
u_t=u_{xxx}+3uu_x,\\[1mm]
v_t=v_{xxx}+3u_xv+3uv_x.
\end{cases}
\end{equation}
We refer to \eqref{eq:local-kdv-ghost} as the local KdV--ghost
system.

The projection \(
    C\longmapsto
    v=\partial_xC
\) identifies ghost representatives that differ by an
\(x\)-independent function. The shift symmetry has been removed. As will be shown in
Section~4.3, the triangular zero-curvature
representation determines the equation for
\(C\) only up to a term \(\Gamma(t)\).
Passing to \(v\) removes this integration ambiguity and yields
the closed local system \eqref{eq:local-kdv-ghost}.

\subsection{Fréchet linearization and tangent interpretation}
\label{subsec:tangent-formulation}

Before turning to the Darboux transformation, it is useful to record a
structural property of the local formulation. After introducing
\(v=C_x\), the second equation of the KdV--ghost system coincides with
the linearized KdV equation along the solution \(u\). In fact, it has the same form as the generator of infinitesimal real symmetries, see \cite{FukuyamaKamimuraKresicJuricMeljanac2007}. However, being the ghost field a nilpotent one, the BRST symmetry is an exact symmetry of the system.  Although the full local ghost equation has the form of the Grassmann-valued
Fréchet linearization of the KdV flow, its decomposition into Grassmann
components differs from the ordinary real tangent dynamics.

The equations corresponding to the nilpotent components of the Grassmann
expansion differ from the ordinary real linearized dynamics and should be
interpreted within the full Grassmann-valued formulation.

In fact, let us consider a straightforward case. If
\(
u=u_0+u_1\theta_1\theta_2
\)
satisfies
\[
u_t=u_{xxx}+3uu_x,
\]
then
\[
(u_0)_t=(u_0)_{xxx}+3u_0(u_0)_x,
\]
and
\[
(u_1)_t=(u_1)_{xxx}
+3u_0(u_1)_x
+3u_1(u_0)_x.
\]

The real field $u_0$ satisfies the KdV equation, whereas $u_1$ has a
different dynamics. The same occurs with the ghost equation. It may have the
same differential form as the equation for an infinitesimal generator of a
real symmetry of the KdV equation, but its content is different. It describes
a system of Grassmann-valued equations, and the important point is the exact
invariance of the KdV--ghost system under the nonlinear BRST transformation.

The first-order perturbation system of the KdV equation was studied by
Ma and Fuchssteiner \cite{MaFuchssteiner1996}. To see how this system
arises, consider the KdV equation in evolutionary form
\[
D_tu=K[u],
\qquad
K[u]=D_x^3u+3uD_xu,
\]
and introduce an infinitesimal deformation
\(
u_\varepsilon=u+\varepsilon v,
\quad
\varepsilon^2=0.
\)
In the present Grassmann-valued setting, $\varepsilon$ and $v$ are odd,
so that $\varepsilon v$ is even. Expanding the KdV vector field gives
\[
K[u+\varepsilon v]
=
K[u]+\varepsilon K'[u]v,
\]
where $K'[u]$ denotes the Fr\'echet derivative of $K$ at $u$. Explicitly,
\[
K'[u]v
=
D_x^3v+3(D_xu)v+3uD_xv.
\]
Therefore, imposing
\(
D_tu_\varepsilon=K[u_\varepsilon]
\)
and separating the terms independent of $\varepsilon$ from those
linear in $\varepsilon$ yields
\(
D_tu=K[u],
\qquad
D_tv=K'[u]v.
\)

Thus the second equation describes the tangent dynamics along the KdV
flow. More generally, systems of the form
\begin{equation*}
D_tu=K[u],
\quad
D_tv=K'[u]v
\end{equation*}
were studied by Ma as particular integrable couplings \cite{Ma2000}.

The same tangent interpretation can equivalently be formulated in
terms of the KdV residue.

For an arbitrary even field \(w=w(x,t)\), define the KdV residue by
\begin{equation}
\label{eq:kdv-residue}
\mathcal{R}[w]
:=
w_t-w_{xxx}-3ww_x.
\end{equation}
Its formal linearization at \(w\), acting on a variation \(z\), is
\begin{equation}
\label{eq:linearized-kdv-operator}
\mathcal{L}_{w}[z]
:=
z_t-z_{xxx}-3w_xz-3wz_x.
\end{equation}
The local KdV--ghost system \eqref{eq:local-kdv-ghost} can therefore be
written as
\begin{equation}
\label{eq:local-system-residue}
\mathcal{R}[u]=0,
\qquad
\mathcal{L}_{u}[v]=0.
\end{equation}

To verify this interpretation directly, let \(\varepsilon\) be an
independent odd nilpotent parameter, \(\varepsilon^2=0\), and let \(h\)
be an odd variation. Since \(\varepsilon h\) is even, one obtains
\begin{align*}
\mathcal{R}[u+\varepsilon h]
&=
\mathcal{R}[u]
+
\varepsilon
\left(
h_t-h_{xxx}-3u_xh-3uh_x
\right)
\nonumber\\
&=
\mathcal{R}[u]
+
\varepsilon\mathcal{L}_{u}[h].
\label{eq:frechet-expansion}
\end{align*}
Hence \(\mathcal{L}_{u}\) is the formal linearization of the KdV
residue at \(u\), and the ghost field \(v\) satisfies the linearized
KdV equation along the solution \(u\).

The relation between this tangent formulation, perturbation systems,
and integrable couplings will be discussed in
Subsection~7.1.

\subsection{From the differential spectral problem to the triangular ZCR}
\label{subsec:triangular-zcr}

Consider the extended spectral problem
\begin{equation}
\label{eq:extended-spectral-problem}
\mathscr{L}_{\varepsilon}\Psi=\lambda\Psi,
\qquad
\Psi_t=\mathscr{M}_{\varepsilon}\Psi,
\end{equation}
where \(\lambda\in\mathbb{R}\) is an even spectral parameter. We take \(\Psi\) to be even, with \(\Psi_0\) even and
\(\Psi_1\) odd. Hence the product \(\varepsilon\Psi_1\) is even,
consistently with the parity of the extended spectral problem. We expand the
eigenfunction in the nilpotent parameter as
\begin{equation}
\label{eq:nilpotent-wavefunction-expansion}
\Psi=\Psi_0+\varepsilon\Psi_1.
\end{equation}

Substitution of
\eqref{eq:nilpotent-wavefunction-expansion} into the spatial
spectral equation in \eqref{eq:extended-spectral-problem}, followed
by separation of the terms independent of \(\varepsilon\) and those
linear in \(\varepsilon\), gives
\begin{subequations}
\label{eq:spatial-component-system}
\begin{align}
\Psi_{0,xx}
&=
\left(
\lambda-\frac{1}{2}u
\right)\Psi_0,
\label{eq:psi0-spatial}
\\
\Psi_{1,xx}
&=
\left(
\lambda-\frac{1}{2}u
\right)\Psi_1
-C_x\Psi_{0,x}
-\frac{1}{2}C_{xx}\Psi_0.
\label{eq:psi1-spatial}
\end{align}
\end{subequations}

Similarly, the temporal equation in
\eqref{eq:extended-spectral-problem} yields
\begin{subequations}
\label{eq:temporal-component-system}
\begin{align}
\Psi_{0,t}
&=
-\frac{1}{2}u_x\Psi_0
+(4\lambda+u)\Psi_{0,x},
\label{eq:psi0-temporal}
\\
\Psi_{1,t}
&=
\left[
2\left(
\lambda-\frac{1}{2}u
\right)C_x
-\frac{1}{2}C_{xxx}
\right]\Psi_0
-\frac{1}{2}u_x\Psi_1
+(4\lambda+u)\Psi_{1,x}.
\label{eq:psi1-temporal}
\end{align}
\end{subequations}
In obtaining \eqref{eq:temporal-component-system}, the spatial
equations \eqref{eq:spatial-component-system} have been used to
eliminate higher \(x\)-derivatives of \(\Psi_0\) and \(\Psi_1\).

We now introduce the vector
\begin{equation*}
\label{eq:extended-wave-vector}
\boldsymbol{\Psi}=
\begin{pmatrix}
\Psi_0\\
\Psi_{0,x}\\
\Psi_1\\
\Psi_{1,x}
\end{pmatrix}.
\end{equation*}
Equations \eqref{eq:spatial-component-system} and
\eqref{eq:temporal-component-system} can then be written as the
first-order linear system
\begin{equation}
\label{eq:triangular-linear-system}
\boldsymbol{\Psi}_x=\mathcal{U}\boldsymbol{\Psi},
\qquad
\boldsymbol{\Psi}_t=\mathcal{V}\boldsymbol{\Psi},
\end{equation}
where the matrices \(\mathcal{U}\) and \(\mathcal{V}\) have the
triangular block form
\begin{equation}
\label{eq:triangular-block-form}
\mathcal{U}
=
\begin{pmatrix}
U_0 & 0\\
U_1 & U_0
\end{pmatrix},
\qquad
\mathcal{V}
=
\begin{pmatrix}
V_0 & 0\\
V_1 & V_0
\end{pmatrix}.
\end{equation}

The spatial blocks are
\begin{equation}
\label{eq:spatial-zcr-blocks}
U_0=
\begin{pmatrix}
0 & 1\\[1mm]
\displaystyle\lambda-\frac{1}{2}u & 0
\end{pmatrix},
\qquad
U_1=
\begin{pmatrix}
0 & 0\\[1mm]
\displaystyle-\frac{1}{2}C_{xx} & -C_x
\end{pmatrix}.
\end{equation}
The temporal blocks are
\begin{equation}
\label{eq:temporal-zcr-blocks}
V_0=
\begin{pmatrix}
-\dfrac{1}{2}u_x
&
4\lambda+u
\\[2mm]
4\lambda^2-\lambda u-\dfrac{1}{2}u^2-\dfrac{1}{2}u_{xx}
&
\dfrac{1}{2}u_x
\end{pmatrix},
\end{equation}
and
\begin{equation}
\label{eq:ghost-temporal-zcr-block}
V_1=
\begin{pmatrix}
\displaystyle
2\left(
\lambda-\frac{1}{2}u
\right)C_x
-\frac{1}{2}C_{xxx}
&
0
\\[3mm]
\displaystyle
-u_xC_x
-\frac{3}{2}uC_{xx}
-\frac{1}{2}C_{xxxx}
&
\displaystyle
-(2\lambda+2u)C_x
-\frac{1}{2}C_{xxx}
\end{pmatrix}.
\end{equation}

The diagonal blocks \((U_0,V_0)\) constitute the standard
first-order matrix representation of the KdV spectral problem,
whereas the lower-left blocks \((U_1,V_1)\) contain the ghost
contribution. The compatibility condition of
\eqref{eq:triangular-linear-system} will be analyzed in the next
subsection.

\subsection{Zero-curvature equations and closure in the local variable}
\label{subsec:zcr-closure}

The compatibility condition for the first-order system~(29) is
\begin{equation*}
\boldsymbol{\Psi}_{xt}-\boldsymbol{\Psi}_{tx}=0.
\end{equation*}
Using~(29), this condition is equivalent to the zero-curvature equation
\begin{equation}
\label{eq:zero-curvature}
\mathcal{Z}
:=
\mathcal{U}_t-\mathcal{V}_x
+[\mathcal{U},\mathcal{V}]
=0.
\end{equation}

Since \(\mathcal{U}\) and \(\mathcal{V}\) have the triangular block
form~(30), the curvature matrix has the same block structure,
\begin{equation}
\label{eq:curvature-block-form}
\mathcal{Z}=
\begin{pmatrix}
Z_0&0\\
Z_1&Z_0
\end{pmatrix},
\end{equation}
where
\begin{align}
Z_0&=
U_{0,t}-V_{0,x}+[U_0,V_0],
\label{eq:Z0-definition}\\
Z_1&=
U_{1,t}-V_{1,x}
+U_1V_0-V_0U_1
+U_0V_1-V_1U_0.
\label{eq:Z1-definition}
\end{align}

For the diagonal block, direct substitution of
\(\mathcal{U}_0\) and \(\mathcal{V}_0\) gives
\begin{equation}
\label{eq:Z0-residue}
Z_0
=
\begin{pmatrix}
0&0\\[1mm]
-\dfrac12\mathcal{R}[u]&0
\end{pmatrix},
\end{equation}
where \(\mathcal{R}\) is the KdV residue defined \eqref{eq:kdv-residue}.

For the lower-left block, recall the representative ghost residue
\begin{equation}
\label{eq:ghost-residue}
\Theta[u,C]
:=
C_t-C_{xxx}-3uC_x.
\end{equation}
A direct calculation gives
\begin{equation*}
\label{eq:Z1-Theta}
Z_1
=
\begin{pmatrix}
0&0\\[1mm]
-\dfrac12D_x^2\Theta[u,C]&-D_x\Theta[u,C]
\end{pmatrix}.
\end{equation*}

Since \(v=C_x\), differentiation of~\eqref{eq:ghost-residue}
with respect to \(x\) gives
\begin{equation}
\label{eq:ghost-residue-linearization}
D_x\Theta[u,C]
=
v_t-v_{xxx}-3u_xv-3uv_x
=
\mathcal{L}_u[v].
\end{equation}
where \(\mathcal{L}_u\) is the linearized KdV operator defined in
Subsection~\ref{subsec:tangent-formulation}. Therefore,
\begin{equation}
\label{eq:Z1-local}
Z_1=
\begin{pmatrix}
0&0\\[1mm]
-\dfrac12 D_x\mathcal{L}_u[v]&
-\mathcal{L}_u[v]
\end{pmatrix}.
\end{equation}

We can now state the compatibility result directly in terms of the
local fields.

\begin{teorema}[Triangular zero-curvature representation]
\label{thm:triangular-zcr}
Let $U$ and $V$ be the matrices appearing in (29) and defined by
(30)--(33), and let $v=C_x$.
Then the zero-curvature condition \eqref{eq:zero-curvature} is equivalent to
\begin{equation}\label{RL=0}
\mathcal{R}[u]=0,
\qquad
\mathcal{L}_u[v]=0.
\end{equation}
Hence the triangular zero-curvature condition is equivalent to the
local KdV--ghost system.
\end{teorema}

\begin{proof}
By \eqref{eq:curvature-block-form}, the zero-curvature condition is
equivalent to \(
Z_0=0,
\quad
Z_1=0.
\) Equation~\eqref{eq:Z0-residue} shows that
\(
Z_0=0 \) if and only if \(
\mathcal{R}[u]=0.
\) Likewise, from \eqref{eq:Z1-local},
\(
Z_1=0 \) if and only if \(
\mathcal{L}_u[v]=0,
\)
because the lower-right entry of \(Z_1\) is
\(-\mathcal{L}_u[v]\).
Therefore the zero-curvature condition is equivalent to \eqref{RL=0}
which is precisely the local KdV--ghost system.
\end{proof}

At the representative level, \(
D_x\Theta[u,C]=0,
\)
and hence
\(
\Theta[u,C]=\Gamma(t),
\)
where \(\Gamma(t)\) is odd and independent of \(x\). Thus
\[
C_t=C_{xxx}+3uC_x+\Gamma(t).
\]
Passing to \(v=C_x\) eliminates \(\Gamma(t)\) and yields the local
KdV--ghost system \eqref{eq:local-kdv-ghost}.

Under the Schwartz-class assumptions of Subsection~2.2,
\(\Theta[u,C]\in\mathcal{S}(\mathbb{R})\) for each fixed \(t\).
Since \(D_x\Theta[u,C]=0\), it follows that \(
\Theta[u,C]=0,
\)
and therefore \(\Gamma(t)=0\).

\section{Nilpotent Darboux transformation}
\label{sec:darboux}

We now construct a nilpotent lift of the scalar KdV Darboux
transformation and prove that it closes on the local KdV--ghost
system \eqref{eq:local-kdv-ghost}. The construction is based on the
Riccati formulation of the Schrödinger spectral problem associated
with KdV.

The proof does not require the explicit construction of a matrix
dressing transformation for the full triangular zero-curvature
representation. Instead, closure on the local system will be
established through factorization identities for the KdV residue
and its Fréchet derivative.

\subsection{Scalar KdV Darboux transformation}
\label{subsec:scalar-darboux}

Let \(\psi=\psi(x,t;\lambda)\) denote a scalar eigenfunction of the
bosonic KdV Lax pair,
\begin{equation*}
\mathscr{L}_0\psi=\lambda\psi,
\qquad
\psi_t=\mathscr{M}_0\psi,
\end{equation*}
where \(\mathscr{L}_0\) and \(\mathscr{M}_0\) are the operators defined in~\eqref{eq:bosonic-lax-pair}.
The spatial equation is equivalently
\begin{equation}\label{SESCHR}
\psi_{xx}
=
\left(\lambda-\frac12u\right)\psi.
\end{equation}

Fix an even spectral value
\(
\lambda=\alpha\in\mathbb{R},
\)
and let \(\varphi\) be a nonzero eigenfunction of
\(\mathscr{L}_0\) at \(\lambda=\alpha\). On any open set on which
\(\varphi\neq0\), define the Riccati variable
\begin{equation}
\label{eq:riccati-variable}
p
:=
\frac{\varphi_x}{\varphi}.
\end{equation}
Differentiating \eqref{eq:riccati-variable} and using
\eqref{SESCHR} gives
\begin{equation*}
\label{eq:riccati-spatial-equation}
p_x
=
\alpha-\frac{1}{2}u-p^2.
\end{equation*}
Equivalently, the potential \(u\) admits the Miura-type
representation
\begin{equation}
\label{eq:u-miura-minus}
u
=
2\alpha-2p^2-2p_x.
\end{equation}

The classical Darboux transformation for the KdV spectral problem is
well known; see, for example, \cite{MatveevSalle1991,GuHuZhou2005}.
The Darboux operator associated with $\phi$ is
\(
D_0=D_x-p.
\)

It intertwines the Schrödinger operators according to \(
\widetilde{\mathscr{L}}_0\mathcal{D}_0
=
\mathcal{D}_0\mathscr{L}_0,
\)
where \(
\widetilde{\mathscr{L}}_0
=
D_x^2+\dfrac{1}{2}\widetilde{u}\,I.
\) The transformed potential is
\begin{equation*}
\label{eq:scalar-darboux-potential}
\widetilde{u}
=
u+4p_x.
\end{equation*}
Combining \eqref{eq:u-miura-minus} with the Darboux transformation above, we obtain
\begin{equation*}
\label{eq:u-miura-plus}
\widetilde{u}
=
2\alpha-2p^2+2p_x.
\end{equation*}

The temporal part of the KdV Lax pair evaluated at
\(\lambda=\alpha\) yields
\begin{equation}
\label{eq:riccati-temporal-equation}
p_t
=
\left[
-\frac{1}{2}u_x
+
(4\alpha+u)p
\right]_x.
\end{equation}
Using \eqref{eq:u-miura-minus}, this equation can be written solely
in terms of \(p\) as
\begin{equation}
\label{eq:p-evolution-equation}
E_p
:=
p_t-p_{xxx}
-6\left(\alpha-p^2\right)p_x
=
0.
\end{equation}

\subsection{Nilpotent lift of the Riccati system}
\label{subsec:nilpotent-lift}

We now lift the scalar Riccati system to the local ghost sector.
Introduce the nilpotent displacements
\begin{equation}
\label{eq:nilpotent-displacements}
u_{\varepsilon}
=
u+\varepsilon v,
\qquad
p_{\varepsilon}
=
p+\varepsilon\eta,
\qquad
\varepsilon^2=0.
\end{equation}
Here \(u\) and \(p\) are even, whereas \(v\) and \(\eta\) are odd.
Consequently, both \(u_{\varepsilon}\) and \(p_{\varepsilon}\) are
even.

Substituting \eqref{eq:nilpotent-displacements} into the Riccati
equation
\[
(p_{\varepsilon})_x
=
\alpha-\frac{1}{2}u_{\varepsilon}
-p_{\varepsilon}^2
\]
and separating the terms independent of \(\varepsilon\) from those
linear in \(\varepsilon\), we recover
\begin{equation}
\label{eq:riccati-system-nilpotent}
p_x
=
\alpha-\frac{1}{2}u-p^2, \quad
\eta_x
=
-\frac{1}{2}v-2p\eta.
\end{equation}
The second equation in
\eqref{eq:riccati-system-nilpotent} is the Fréchet linearization of
the scalar Riccati equation in the direction \((v,\eta)\).

The scalar Darboux transformation
\[
\widetilde{u}_{\varepsilon}
=
u_{\varepsilon}
+
4(p_{\varepsilon})_x
\]
has the nilpotent expansion
\(
\widetilde{u}_{\varepsilon}
=
\widetilde{u}
+
\varepsilon\widetilde{v}.
\)
Therefore,
\begin{equation}
\label{eq:nilpotent-darboux-fields}
\widetilde{u}
=
u+4p_x,
\qquad
\widetilde{v}
=
v+4\eta_x.
\end{equation}

Solving the second equation of
\eqref{eq:riccati-system-nilpotent} for \(v\), and then using
\eqref{eq:nilpotent-darboux-fields}, gives the equivalent
Miura-type relations
\begin{equation*}
\label{eq:ghost-miura-relations}
v
=
-2\eta_x-4p\eta,
\qquad
\widetilde{v}
=
2\eta_x-4p\eta.
\end{equation*}

We next linearize the temporal Riccati equation
\eqref{eq:riccati-temporal-equation}. Substitution of
\eqref{eq:nilpotent-displacements} into
\[
(p_{\varepsilon})_t
=
\left[
-\frac{1}{2}(u_{\varepsilon})_x
+
(4\alpha+u_{\varepsilon})p_{\varepsilon}
\right]_x
\]
yields, at first order in \(\varepsilon\),
\begin{equation}
\label{eq:eta-temporal-equation}
\eta_t
=
\left[
-\frac{1}{2}v_x
+
vp
+
(4\alpha+u)\eta
\right]_x.
\end{equation}

Equivalently, the Fréchet linearization of
\eqref{eq:p-evolution-equation} in the direction \(\eta\) is
\begin{equation*}
\label{eq:E-eta-definition}
E_{\eta}
:=
\eta_t-\eta_{xxx}
-6\left(\alpha-p^2\right)\eta_x
+12pp_x\eta.
\end{equation*}
Equations
\eqref{eq:riccati-system-nilpotent} and
\eqref{eq:eta-temporal-equation} imply \(E_{\eta}=0.
\)

\subsection{Factorization identities}
\label{subsec:factorization}

The closure of the nilpotent Darboux transformation on the local
KdV--ghost system will follow from a set of off-shell factorization
identities. For this purpose, we now regard \(p\) and \(\eta\) as
arbitrary sufficiently smooth fields, with \(p\) even and \(\eta\)
odd, and define

\begin{align}
u
&=
2\alpha-2p^2-2p_x,
&
\widetilde{u}
&=
2\alpha-2p^2+2p_x,
\label{eq:factorization-u-fields}
\\
v
&=
-2\eta_x-4p\eta,
&
\widetilde{v}
&=
2\eta_x-4p\eta.
\label{eq:factorization-v-fields}
\end{align}

Recall the KdV residue and its Fréchet derivative, defined in
\eqref{eq:kdv-residue} and
\eqref{eq:linearized-kdv-operator}. We also recall
\begin{align}
E_p
&=
p_t-p_{xxx}
-6\left(\alpha-p^2\right)p_x,
\label{eq:factorization-Ep}
\\
E_{\eta}
&=
\eta_t-\eta_{xxx}
-6\left(\alpha-p^2\right)\eta_x
+12pp_x\eta.
\label{eq:factorization-Eeta}
\end{align}

\begin{proposicion}[Factorization identities]
\label{prop:factorization-identities}
For arbitrary sufficiently smooth \(p\) and \(\eta\), the fields
defined by
\eqref{eq:factorization-u-fields} and
\eqref{eq:factorization-v-fields} satisfy
\begin{subequations}
\label{eq:factorization-identities}
\begin{align}
\mathcal{R}[u]
&=
-2\left(D_x+2p\right)E_p,
\label{eq:factorization-R-u}
\\
\mathcal{R}[\widetilde{u}]
&=
2\left(D_x-2p\right)E_p,
\label{eq:factorization-R-utilde}
\\
\mathcal{L}_u[v]
&=
-2\left[
\left(D_x+2p\right)E_{\eta}
+2\eta E_p
\right],
\label{eq:factorization-L-u}
\\
\mathcal{L}_{\widetilde{u}}[\widetilde{v}]
&=
2\left[
\left(D_x-2p\right)E_{\eta}
-2\eta E_p
\right].
\label{eq:factorization-L-utilde}
\end{align}
\end{subequations}
\end{proposicion}

\begin{proof}
We first prove the two bosonic identities. From
\[
u=2\alpha-2p^2-2p_x,
\]
we obtain
\[
u_t=
-4pp_t-2p_{xt}, \qquad
u_x
=
-4pp_x-2p_{xx}, \qquad
u_{xxx}
=
-12p_xp_{xx}-4pp_{xxx}-2p_{xxxx}.
\]

Substitution into
\(
\mathcal{R}[u]
\)
and collection of terms gives
\[
\mathcal{R}[u]
=
-2\partial_x E_p-4pE_p
=
-2(D_x+2p)E_p,
\]
which proves~\eqref{eq:factorization-R-u}.

Similarly, substituting
\[
\widetilde{u}
=
2\alpha-2p^2+2p_x
\]
into the KdV residue gives
\[
\mathcal{R}[\widetilde{u}]
=
2\left(D_x-2p\right)E_p,
\]
which proves \eqref{eq:factorization-R-utilde}. Equivalently, this
second identity follows from the first one under the involution
\(
p\longmapsto -p.
\) Indeed, under \(p\mapsto -p\) one has
\(u\mapsto\widetilde u\) and \(E_p\mapsto -E_p\).

We now derive the ghost identities by nilpotent linearization. Let
\(
p_{\varepsilon}
=
p+\varepsilon\eta,
\quad
\varepsilon^2=0,
\)
where \(\varepsilon\) and \(\eta\) are odd, so that
\(p_{\varepsilon}\) is even. The corresponding fields satisfy
\begin{align*}
u_{\varepsilon}
&=
2\alpha-2p_{\varepsilon}^2
-2(p_{\varepsilon})_x
=
u+\varepsilon v,
\\
\widetilde{u}_{\varepsilon}
&=
2\alpha-2p_{\varepsilon}^2
+2(p_{\varepsilon})_x
=
\widetilde{u}
+\varepsilon\widetilde{v}.
\end{align*}
Moreover,
\begin{equation}
\label{eq:Ep-nilpotent-expansion}
E_{p_{\varepsilon}}
=
E_p+\varepsilon E_{\eta}.
\end{equation}

Since \(p_\varepsilon\) is even, the algebraic identity~(52a)
applies to \(p_\varepsilon\). In fact,
\[
\mathcal{R}[u_{\varepsilon}]
=
-2
\left(
D_x+2p_{\varepsilon}
\right)
E_{p_{\varepsilon}}.
\]
Using \(
\mathcal{R}[u+\varepsilon v]
=
\mathcal{R}[u]
+\varepsilon\mathcal{L}_u[v]
\) and \eqref{eq:Ep-nilpotent-expansion}, the coefficient of
\(\varepsilon\) yields
\[
\mathcal{L}_u[v]
=
-2\left[
\left(D_x+2p\right)E_{\eta}
+2\eta E_p
\right].
\]
This proves \eqref{eq:factorization-L-u}.

Likewise, applying \eqref{eq:factorization-R-utilde} to
\(p_{\varepsilon}\) gives
\[
\mathcal{R}[\widetilde{u}_{\varepsilon}]
=
2
\left(
D_x-2p_{\varepsilon}
\right)
E_{p_{\varepsilon}}.
\]
Taking the coefficient of \(\varepsilon\), we obtain
\[
\mathcal{L}_{\widetilde{u}}[\widetilde{v}]
=
2\left[
\left(D_x-2p\right)E_{\eta}
-2\eta E_p
\right],
\]
which proves \eqref{eq:factorization-L-utilde}.
\end{proof}

The identities in
\eqref{eq:factorization-identities} hold without assuming
\(E_p=0\) or \(E_{\eta}=0\). In particular, when the auxiliary
Riccati fields satisfy \(
E_p=0,
\quad
E_{\eta}=0,
\) they imply simultaneously
\(
\mathcal{R}[u]
=
\mathcal{R}[\widetilde{u}]
=
0
\)
and
\(
\mathcal{L}_u[v]
=
\mathcal{L}_{\widetilde{u}}[\widetilde{v}]
=
0.
\)

These identities will be used below to prove the closure of the
nilpotent Darboux transformation.

\subsection{Compatibility of the auxiliary system}
\label{subsec:compatibility}

The auxiliary odd field \(\eta\) is governed by the spatial and
temporal equations
\begin{subequations}
\label{eq:eta-auxiliary-system}
\begin{align}
\eta_x
&=
-\frac{1}{2}v-2p\eta,
\label{eq:eta-auxiliary-spatial}
\\
\eta_t
&=
\left[
-\frac{1}{2}v_x
+vp
+(4\alpha+u)\eta
\right]_x.
\label{eq:eta-auxiliary-temporal}
\end{align}
\end{subequations}
Before stating the Darboux theorem, we verify that this overdetermined
system is compatible whenever the seed fields satisfy the local
KdV--ghost equations.

\begin{lema}[Compatibility of the auxiliary ghost system]
\label{lem:auxiliary-compatibility}
Let \((u,v)\) be a sufficiently smooth solution of the local
KdV--ghost system \eqref{eq:local-system-residue} and let \(p\) satisfy
\begin{subequations}
\label{eq:p-auxiliary-system}
\begin{align}
p_x
&=
\alpha-\frac{1}{2}u-p^2,
\label{eq:p-auxiliary-spatial}
\\
p_t
&=
\left[
-\frac{1}{2}u_x
+(4\alpha+u)p
\right]_x.
\label{eq:p-auxiliary-temporal}
\end{align}
\end{subequations}
Then the auxiliary system
\eqref{eq:eta-auxiliary-system} is compatible:
\begin{equation*}
\label{eq:eta-compatibility}
\partial_t(\eta_x)
=
\partial_x(\eta_t).
\end{equation*}

Moreover, if \(\eta_1\) and \(\eta_2\) are two solutions of
\eqref{eq:eta-auxiliary-system}, then their difference
\(
\rho:=\eta_1-\eta_2
\)
satisfies the homogeneous system
\begin{subequations}
\label{eq:eta-homogeneous-system}
\begin{align}
\rho_x
&=
-2p\rho,
\label{eq:rho-spatial}
\\
\rho_t
&=
\left[
(4\alpha+u)\rho
\right]_x.
\label{eq:rho-temporal}
\end{align}
\end{subequations}
\end{lema}

\begin{proof}
Introduce the functions
\begin{align*}
a[u,p]
&:=
\alpha-\frac{1}{2}u-p^2, \qquad
b[u,p]:=
-\frac{1}{2}u_x+(4\alpha+u)p.
\label{eq:b-functional}
\end{align*}
Then the auxiliary equations for \(p\) can be written as
\begin{equation*}
\label{eq:p-system-ab}
p_x=a[u,p],
\qquad
p_t=D_xb[u,p].
\end{equation*}

Using the auxiliary equations for \(p\), a direct calculation gives
\begin{equation}
\label{eq:p-compatibility-residue}
\partial_t a[u,p]
-
D_x^2 b[u,p]
=
-\frac{1}{2}\mathcal{R}[u].
\end{equation}
Indeed,
\begin{align*}
\partial_t a[u,p]
-
D_x^2 b[u,p]
&=
-\frac{1}{2}u_t
-2pp_t
+\frac{1}{2}u_{xxx}
-u_{xx}p
-2u_xp_x
-(4\alpha+u)p_{xx}.
\end{align*}
From the auxiliary system,
\[
p_t
=
-\frac{1}{2}u_{xx}
+u_xp
+(4\alpha+u)p_x,
\]
whereas differentiation of the spatial Riccati equation gives
\[
p_{xx}
=
-\frac{1}{2}u_x
-2pp_x.
\]
Substitution of these expressions yields
\[
\partial_t a[u,p]
-
D_x^2 b[u,p]
=
-\frac{1}{2}
\left(
u_t-u_{xxx}-3uu_x
\right)
=
-\frac{1}{2}\mathcal{R}[u].
\]

Consequently, the compatibility of the auxiliary system for \(p\)
follows from \(\mathcal{R}[u]=0\).

Consider now the nilpotent deformation
\(
u_{\varepsilon}
=
u+\varepsilon v,
\quad
p_{\varepsilon}
=
p+\varepsilon\eta,
\quad
\varepsilon^2=0.
\)
The Fréchet derivatives of \(a\) and \(b\) in the direction
\((v,\eta)\) are
\begin{align*}
D a[u,p](v,\eta)
&=
-\frac{1}{2}v-2p\eta,
\\
D b[u,p](v,\eta)
&=
-\frac{1}{2}v_x
+vp
+(4\alpha+u)\eta.
\end{align*}
Thus,
\[
\eta_x
=
D a[u,p](v,\eta),
\qquad
\eta_t
=
D_xD b[u,p](v,\eta).
\]

Taking the Fréchet derivative of
\eqref{eq:p-compatibility-residue} in the direction \((v,\eta)\)
gives
\[
\partial_t
\left(
D a[u,p](v,\eta)
\right)
-
D_x^2
\left(
D b[u,p](v,\eta)
\right)
=
-\frac{1}{2}\mathcal{L}_u[v].
\]
Using the auxiliary equations for \(\eta\), this becomes
\[
\partial_t(\eta_x)
-
D_x(\eta_t)
=
-\frac{1}{2}\mathcal{L}_u[v].
\]
Since the seed fields satisfy
\(
\mathcal{L}_u[v]=0,
\)
we conclude that
\(
\partial_t(\eta_x)
=
\partial_x(\eta_t).
\)
Finally, let \(\eta_1\) and \(\eta_2\) be two solutions of
\eqref{eq:eta-auxiliary-system}. Subtracting their spatial
equations gives
\[
(\eta_1-\eta_2)_x
=
-2p(\eta_1-\eta_2),
\]
whereas subtracting their temporal equations gives
\[
(\eta_1-\eta_2)_t
=
\left[
(4\alpha+u)(\eta_1-\eta_2)
\right]_x.
\]
Therefore, setting
\(
\rho:=\eta_1-\eta_2,
\)
we obtain
\[
\rho_x=-2p\rho,
\qquad
\rho_t=
\left[
(4\alpha+u)\rho
\right]_x,
\]
which proves \eqref{eq:eta-homogeneous-system}.
\end{proof}

\subsection{The nilpotent Darboux theorem}
\label{subsec:main-theorem}

We can now summarize the preceding construction in the following
result.

\begin{teorema}[Nilpotent Darboux transformation]
\label{thm:nilpotent-darboux-transformation}
Let \((u,v)\) be a sufficiently smooth solution of the local
KdV--ghost system \eqref{eq:local-system-residue}. Fix an even parameter \(\alpha\in\mathbb{R}\). Suppose that the even
field \(p\) and the odd field \(\eta\) satisfy
\begin{subequations}
\label{eq:auxiliary-system-darboux-theorem}
\begin{align}
p_x
&=
\alpha-\frac{1}{2}u-p^2,
\label{eq:p-spatial-darboux-theorem}
\\
p_t
&=
\left[
-\frac{1}{2}u_x+(4\alpha+u)p
\right]_x,
\label{eq:p-temporal-darboux-theorem}
\\
\eta_x
&=
-\frac{1}{2}v-2p\eta,
\label{eq:eta-spatial-darboux-theorem}
\\
\eta_t
&=
\left[
-\frac{1}{2}v_x+vp+(4\alpha+u)\eta
\right]_x.
\label{eq:eta-temporal-darboux-theorem}
\end{align}
\end{subequations}

Define \((\widetilde u,\widetilde v)\) by~\eqref{eq:nilpotent-darboux-fields}. Then the transformed
pair also satisfies the local KdV--ghost system: \( \mathcal{R}[\widetilde{u}]
=
0,
\quad
\mathcal{L}_{\widetilde{u}}[\widetilde{v}]
=
0.
\)

\end{teorema}

\begin{proof}
The auxiliary equations for \(p\) and \(\eta\) imply \(
E_p=0,
\quad
E_{\eta}=0.
\)
The factorization identities of
Proposition~\ref{prop:factorization-identities} therefore give
\[
\mathcal{R}[\widetilde{u}]
=
2(D_x-2p)E_p
=
0
\]
and
\[
\mathcal{L}_{\widetilde{u}}[\widetilde{v}]
=
2\left[
(D_x-2p)E_{\eta}-2\eta E_p
\right]
=
0.
\]
Hence \((\widetilde{u},\widetilde{v})\) satisfies the local
KdV--ghost system.
\end{proof}

The compatibility of the spatial and temporal equations for
\(\eta\) follows from
Lemma~\ref{lem:auxiliary-compatibility}. Thus the theorem defines
a consistent nilpotent extension of the scalar KdV Darboux
transformation.

The equivalent Miura-type representation is
\begin{equation*}
\label{eq:nilpotent-miura-summary}
\begin{aligned}
u
&=
2\alpha-2p^2-2p_x,
&
\widetilde{u}
&=
2\alpha-2p^2+2p_x,
\\
v
&=
-2\eta_x-4p\eta,
&
\widetilde{v}
&=
2\eta_x-4p\eta.
\end{aligned}
\end{equation*}
These symmetric relations provide the corresponding
Bäcklund-type interpretation of the transformation.

\subsection{Lift to the ghost representative}
\label{subsec:relation-triangular-zcr}

We now relate this local transformation to the original ghost
representative \(C\). Let
\(
\pi(u,C):=(u,C_x)=(u,v)
\)
denote the projection from the representative formulation to the
local system. For a fixed admissible pair of auxiliary fields
\((p,\eta)\), define
\(
\Theta_{p,\eta}(u,v)
:=
(\widetilde{u},\widetilde{v})
\)
and lift this transformation to the representative level by
\(
\widehat{\Theta}_{p,\eta}(u,C)
:=
(\widetilde{u},\widetilde{C}),
\quad
\widetilde{C}:=C+4\eta.
\)
Then
\begin{equation*}
\label{eq:darboux-projection-commutes}
\pi\circ\widehat{\Theta}_{p,\eta}
=
\Theta_{p,\eta}\circ\pi,
\end{equation*}
since
\[
\widetilde{C}_x
=
(C+4\eta)_x
=
C_x+4\eta_x
=
v+4\eta_x
=
\widetilde{v}.
\]
Equivalently, the following diagram commutes:
\[
\scalebox{1.25}{$
\begin{array}{ccc}
(u,C)
&
\xrightarrow{\ \widehat{\Theta}_{p,\eta}\ }
&
(\widetilde{u},\widetilde{C})
\\[3mm]
{\scriptstyle\pi}\big\downarrow
&
&
\big\downarrow{\scriptstyle\pi}
\\[3mm]
(u,v)
&
\xrightarrow{\ \Theta_{p,\eta}\ }
&
(\widetilde{u},\widetilde{v})
\end{array}
$}
\]

The commutative diagram shows that the transformation of the ghost
representative is consistent with the induced transformation of the
local ghost variable. Since
\((\widetilde{u},\widetilde{v})\) satisfies the local KdV--ghost
system and \(\widetilde{v}=\widetilde{C}_x\), one obtains
\begin{equation*}
\label{eq:transformed-representative-ambiguity}
\partial_x
\left(
\widetilde{C}_t
-\widetilde{C}_{xxx}
-3\widetilde{u}\widetilde{C}_x
\right)
=
0.
\end{equation*}
Consequently, on each connected component of the \(x\)-domain,
there exists an odd function
\(\widetilde{\Gamma}(t)\), independent of \(x\), such that
\begin{equation*}
\label{eq:transformed-C-equation}
\widetilde{C}_t
=
\widetilde{C}_{xxx}
+
3\widetilde{u}\widetilde{C}_x
+
\widetilde{\Gamma}(t).
\end{equation*}
This is the same \(x\)-independent ambiguity encountered in the
triangular zero-curvature representation. It disappears after differentiation with respect to \(x\).

\section{Explicit examples}
\label{sec:examples}

In this section, we illustrate the nilpotent Darboux construction developed
above by means of explicit solitonic examples. We first consider the
vacuum seed and show how a single Darboux step generates simultaneously
the standard KdV one-soliton and a nontrivial local ghost field. We then
extend the construction to successive Darboux transformations, where the
bosonic Darboux chain is accompanied by a recursive ghost sector with a
new odd homogeneous contribution at each step. These examples also make
explicit the relation between the local ghost fields and tangent
directions associated with the soliton phase parameters.

\subsection{Vacuum-to-soliton transformation}
\label{subsec:vacuum-to-soliton}

For the vacuum seed, the spatial spectral equation~\eqref{SESCHR}, evaluated at the
Darboux value \(\lambda=\alpha\), reduces to \(
\phi_{xx}=\alpha\phi.
\)
To obtain the regular hyperbolic Darboux seed used below, we take
\(\alpha>0\) and introduce
\(
\kappa:=\sqrt{\alpha}>0,
\qquad\text{so that}\qquad
\alpha=\kappa^2.
\)
The spatial and temporal eigenfunction equations therefore become
\(
\phi_{xx}=\kappa^2\phi,
\quad
\phi_t=4\kappa^2\phi_x.
\)
Both equations are satisfied by
\(
\phi=\cosh\xi,
\quad
\xi=\kappa x+4\kappa^3t+\xi_0.
\)

Since $\phi=\cosh\xi$ does not vanish, the Riccati field is well defined,
\[
p=\frac{\phi_x}{\phi}=\kappa\tanh\xi.
\]

We begin with the vacuum seed
$u=v=0$.
At the level of the ghost representative, we also take \(C=0\).

In particular,
\[
p_x
=
\kappa^2\operatorname{sech}^2\xi,
\qquad
p_t
=
4\kappa^4\operatorname{sech}^2\xi
=
4\kappa^2p_x,
\]
so \(p\) satisfies both auxiliary equations for the vacuum seed.

Although the seed ghost field vanishes, the auxiliary equation for
\(\eta\) admits nontrivial homogeneous solutions. Indeed, when
\(v=0\), the odd auxiliary system becomes
\begin{equation}
\label{eq:vacuum-eta-system}
\eta_x=-2p\eta,
\qquad
\eta_t=4\kappa^2\eta_x.
\end{equation}
Let \(\zeta\) be an \(x\)- and \(t\)-independent odd element of the
underlying Grassmann algebra. Then
\begin{equation*}
\label{eq:vacuum-eta-solution}
\eta
=
\kappa^2\zeta\operatorname{sech}^2\xi
\end{equation*}
solves \eqref{eq:vacuum-eta-system}, since
\[
\eta_x
=
-2\kappa^3\zeta
\operatorname{sech}^2\xi\tanh\xi
=
-2p\eta
\]
and
\[
\eta_t
=
-8\kappa^5\zeta
\operatorname{sech}^2\xi\tanh\xi
=
4\kappa^2\eta_x.
\]

Applying the nilpotent Darboux transformation gives
\begin{equation*}
\label{eq:vacuum-transformed-fields}
\widetilde{u}
=
u+4p_x
=
4\kappa^2\operatorname{sech}^2\xi,
\end{equation*}
and
\begin{equation*}
\label{eq:vacuum-transformed-ghost}
\widetilde{v}
=
v+4\eta_x
=
-8\kappa^3\zeta
\operatorname{sech}^2\xi\tanh\xi.
\end{equation*}
At the representative level, the transformation constructed in
Subsection~5.6 gives
\begin{equation*}
\label{eq:vacuum-transformed-representative}
\widetilde{C}
=
C+4\eta
=
4\kappa^2\zeta\operatorname{sech}^2\xi.
\end{equation*}
Consequently,
\(
\widetilde{C}
=
\zeta\widetilde{u},
\quad
\widetilde{v}
=
\widetilde{C}_x
=
\zeta\widetilde{u}_x.
\) Introducing \(
c=4\kappa^2,
\quad
\xi_0=\kappa x_0,
\)
the phase and transformed fields can be written as
\begin{equation*}
\label{eq:vacuum-soliton-c-parametrization}
\xi
=
\frac{\sqrt{c}}{2}
\left(
x+ct+x_0
\right),
\end{equation*}
and \(
\widetilde{u}
=
c\operatorname{sech}^2\xi.
\)

Writing \(\zeta=\mu\vartheta\), where
\(\mu\in\mathbb{R}\) and \(\vartheta\) is a fixed odd Grassmann
generator, one recovers the family
\(
\widetilde{C}
=
\mu\widetilde{u}\,\vartheta.
\)

The relation between the homogeneous auxiliary field and the soliton
phase can be made explicit. Since $\xi_0=\kappa x_0$, one has
\[
\frac{\partial p}{\partial x_0}
 = \kappa^2\sech^2\xi,
\]
and hence
\[
\eta=\zeta\,\frac{\partial p}{\partial x_0}.
\]
Consequently,
\[
\widetilde v
 = 4\eta_x
 = \zeta\,\frac{\partial \widetilde u}{\partial x_0}
 = \zeta\,\widetilde u_x.
\]
Thus, already at the one-soliton level, the homogeneous odd freedom in
the auxiliary Riccati system is identified with the tangent direction
generated by a variation of the soliton phase. Since $\widetilde u$
satisfies the KdV equation,
\[
L_{\widetilde u}[\widetilde u_x]
 = D_xR[\widetilde u]=0,
\]
and therefore
\[
L_{\widetilde u}[\widetilde v]=0.
\]

Likewise, since $\widetilde{C}=\zeta\widetilde{u}$,
\[
\widetilde{C}_t
-
\widetilde{C}_{xxx}
-
3\widetilde{u}\,\widetilde{C}_x
=
\zeta
\left(
\widetilde{u}_t
-
\widetilde{u}_{xxx}
-
3\widetilde{u}\,\widetilde{u}_x
\right)
=
0.
\]
Thus, in this example, the transformed representative satisfies the ghost equation with no $x$-independent remainder,
\(
\widetilde{\Gamma}(t)=0.
\)

\subsection{Successive Darboux transformations}

We now extend the vacuum-to-one-soliton construction of the previous
subsection to a successive Darboux chain. We distinguish throughout
between eigenfunctions of the vacuum spectral problem and the dressed
eigenfunctions that generate the successive Darboux transformations.

Let \(
0<\kappa_1<\kappa_2<\kappa_3<\cdots,
\quad
\xi_j=\kappa_j x+4\kappa_j^3 t+\delta_j,
\)
and denote by $u^{[j]}$ the potential obtained after $j$ Darboux
steps. We set
\[
L^{[j]}:=D_x^2+\frac12 u^{[j]},
\qquad
u^{[0]}=0,
\]
so that $L^{[0]}=D_x^2$.

For each spectral value \(
\alpha_j=\kappa_j^2,
\)
let $\chi_j$ be an eigenfunction of the vacuum Lax pair,
\[
L^{[0]}\chi_j=\kappa_j^2\chi_j,
\qquad
(\chi_j)_t=4\kappa_j^2(\chi_j)_x.
\]
For the regular hyperbolic construction considered below, we choose
\[
\chi_1=\cosh\xi_1,\qquad
\chi_2=\sinh\xi_2,\qquad
\chi_3=\cosh\xi_3,\ldots
\]
Each $\chi_j$ satisfies both equations of the vacuum Lax pair.

The first Darboux transformation is generated by
\(
\phi_1:=\chi_1=\cosh\xi_1.
\)
Since $\phi_1>0$, the associated Riccati field is well defined,
\[
p^{[1]}:=(\log\phi_1)_x
       =\kappa_1\tanh\xi_1.
\]
The first Darboux operator is therefore
\(
D_1:=D_x-p^{[1]},
\)
and the transformed potential is
\[
u^{[1]}
 =u^{[0]}+4p^{[1]}_x
 =4\kappa_1^2\sech^2\xi_1.
\]

We next construct the second Darboux transformation. The function \(
\chi_2=\sinh\xi_2
\)
is an eigenfunction of the vacuum Lax pair at
$\alpha_2=\kappa_2^2$. Since the background after the first Darboux
step is $u^{[1]}$, the corresponding eigenfunction must first be
dressed by $D_1$. We therefore define
\(
\phi_2:=D_1\chi_2.
\)

The spatial intertwining relation
\(
L^{[1]}D_1=D_1L^{[0]}
\)
implies
\[
L^{[1]}\phi_2
 =L^{[1]}D_1\chi_2
 =D_1L^{[0]}\chi_2
 =\kappa_2^2D_1\chi_2
 =\kappa_2^2\phi_2.
\]
Thus $\phi_2$ is an eigenfunction of the spatial operator
$L^{[1]}$ at the spectral value $\alpha_2=\kappa_2^2$.

The temporal equation can be treated directly at the scalar level.
More generally, let $\psi$ satisfy
\[
\psi_{xx}
=
\left(\beta-\frac12 u\right)\psi,
\qquad
\psi_t
=
-\frac12 u_x\psi+(4\beta+u)\psi_x,
\]
and let $p$ satisfy the Riccati equations at the spectral value
$\alpha$,
\[
p_x=\alpha-\frac12u-p^2,
\qquad
p_t=
\left[
-\frac12u_x+(4\alpha+u)p
\right]_x.
\]
Define
\[
\widetilde\psi=(D_x-p)\psi,
\qquad
\widetilde u=u+4p_x.
\]
The usual spatial Darboux relation gives
\[
\widetilde\psi_{xx}
=
\left(\beta-\frac12\widetilde u\right)\widetilde\psi.
\]
For the temporal equation, direct differentiation yields
\[
\begin{aligned}
&\widetilde\psi_t
+\frac12\widetilde u_x\widetilde\psi
-(4\beta+\widetilde u)\widetilde\psi_x
\\
&\quad =
\left(2p_{xx}+u_x+4pp_x\right)\psi_x
\\
&\qquad +
\left(
-2pp_{xx}-p_t+4p_x^2+3up_x-\frac12u_{xx}
\right)\psi .
\end{aligned}
\]
Differentiating the spatial Riccati equation gives
\[
p_{xx}=-\frac12u_x-2pp_x,
\]
so that the coefficient of $\psi_x$ vanishes. Using this relation
together with
\[
p_x=\alpha-\frac12u-p^2,
\]
the coefficient of $\psi$ reduces to
\[
-p_t-\frac12u_{xx}+u_xp+(4\alpha+u)p_x,
\]
which vanishes by the temporal Riccati equation. Therefore
\[
\widetilde\psi_t
=
-\frac12\widetilde u_x\widetilde\psi
+
(4\beta+\widetilde u)\widetilde\psi_x.
\]
Hence the scalar Darboux transformation maps a solution of the
complete scalar Lax pair at spectral value $\beta$ on the background
$u$ to a solution at the same spectral value on the transformed
background $\widetilde u$.

Since $\phi_1=\chi_1$ satisfies the complete vacuum Lax pair at
$\alpha_1=\kappa_1^2$, the field
$p^{[1]}=(\log\phi_1)_x$ satisfies the above Riccati equations with
$u=u^{[0]}$ and $\alpha=\kappa_1^2$. Applying the result with
\[
\psi=\chi_2,\qquad
\beta=\kappa_2^2,
\]
we obtain
\[
(\phi_2)_t
=
-\frac12(u^{[1]})_x\phi_2
+
\left(4\kappa_2^2+u^{[1]}\right)(\phi_2)_x.
\]
Together with the spatial equation proved above, this shows that
$\phi_2=D_1\chi_2$ satisfies the complete scalar Lax pair on the
background $u^{[1]}$ at the spectral value $\kappa_2^2$.

Explicitly,
\[
\phi_2
 =(D_x-p^{[1]})\chi_2
 =\kappa_2\cosh\xi_2
  -\kappa_1\tanh\xi_1\sinh\xi_2
 =\frac{W_{12}}{\cosh\xi_1},
\]
where
\[
W_{12}
 :=W(\chi_1,\chi_2)
 =\kappa_2\cosh\xi_1\cosh\xi_2
  -\kappa_1\sinh\xi_1\sinh\xi_2.
\]

The second Darboux transformation is globally regular. Indeed,
\[
W_{12}
 =\cosh\xi_1\cosh\xi_2
   \left(
   \kappa_2-\kappa_1\tanh\xi_1\tanh\xi_2
   \right).
\]
Since $0<\kappa_1<\kappa_2$ and
$|\tanh\xi_1\tanh\xi_2|<1$,
\[
\kappa_2-\kappa_1\tanh\xi_1\tanh\xi_2
>\kappa_2-\kappa_1>0.
\]
Therefore
\(
W_{12}>0,
\quad
\phi_2>0,
\)
for all $(x,t)$.

The second Riccati field is consequently globally well defined:
\[
p^{[2]}
 :=(\log\phi_2)_x
 =(\log W_{12})_x-p^{[1]}.
\]
Defining
\[
D_2:=D_x-p^{[2]},
\]
the second transformed potential is
\[
u^{[2]}
 =u^{[1]}+4p^{[2]}_x.
\]
Since
\[
u^{[1]}=4p^{[1]}_x,
\qquad
p^{[2]}_x
 =(\log W_{12})_{xx}-p^{[1]}_x,
\]
we obtain
\[
u^{[2]}=4\partial_x^2\log W_{12}.
\]

For completeness, the two-step potential can be written explicitly.
From
\[
W_{12}
 =
\kappa_2\cosh\xi_1\cosh\xi_2
-\kappa_1\sinh\xi_1\sinh\xi_2,
\]
one finds
\[
(W_{12})_x
 =
(\kappa_2^2-\kappa_1^2)
\cosh\xi_1\sinh\xi_2,
\]
and
\[
(W_{12})_{xx}
 =
(\kappa_2^2-\kappa_1^2)
\left(
\kappa_1\sinh\xi_1\sinh\xi_2
+\kappa_2\cosh\xi_1\cosh\xi_2
\right).
\]
Using
\[
\partial_x^2\log W_{12}
=
\frac{
W_{12}(W_{12})_{xx}-(W_{12})_x^2
}{
W_{12}^2
},
\]
together with the hyperbolic identities, gives
\[
W_{12}(W_{12})_{xx}-(W_{12})_x^2
=
(\kappa_2^2-\kappa_1^2)
\left(
\kappa_2^2\cosh^2\xi_1
+\kappa_1^2\sinh^2\xi_2
\right).
\]
Hence
\[
u^{[2]}
=
\frac{
4(\kappa_2^2-\kappa_1^2)
\left(
\kappa_2^2\cosh^2\xi_1
+\kappa_1^2\sinh^2\xi_2
\right)
}{
\left(
\kappa_2\cosh\xi_1\cosh\xi_2
-\kappa_1\sinh\xi_1\sinh\xi_2
\right)^2
}.
\]

Thus the first two Darboux steps give the globally regular chain
\[
u^{[0]}
\;\xrightarrow{\;\phi_1=\chi_1\;}\;
u^{[1]}
\;\xrightarrow{\;\phi_2=D_1\chi_2\;}\;
u^{[2]}.
\]

The construction extends recursively. Let $\chi_j$ be a vacuum
eigenfunction at the spectral value
\[
\alpha_j=\kappa_j^2,
\]
satisfying the complete scalar vacuum Lax pair
\[
L^{[0]}\chi_j=\kappa_j^2\chi_j,
\qquad
(\chi_j)_t=4\kappa_j^2(\chi_j)_x.
\]

We proceed recursively, working at each step on a domain on which
the corresponding dressed eigenfunction does not vanish. Set
\[
\phi_1:=\chi_1.
\]
Suppose that $\phi_r$ has been constructed and satisfies the
complete scalar Lax pair on the background $u^{[r-1]}$ at the
spectral value $\kappa_r^2$. Define
\[
p^{[r]}:=(\log\phi_r)_x,
\qquad
D_r:=D_x-p^{[r]},
\]
and
\[
u^{[r]}=u^{[r-1]}+4p^{[r]}_x.
\]
Since $\phi_r$ satisfies the complete scalar Lax pair,
$p^{[r]}$ satisfies
\[
p^{[r]}_x
=
\kappa_r^2-\frac12u^{[r-1]}-(p^{[r]})^2,
\]
and
\[
p^{[r]}_t
=
\left[
-\frac12(u^{[r-1]})_x
+
\left(4\kappa_r^2+u^{[r-1]}\right)p^{[r]}
\right]_x.
\]
Therefore the scalar Darboux property established above implies
that $D_r$ maps every solution of the complete scalar Lax pair on
the background $u^{[r-1]}$, at an arbitrary spectral value
$\beta$, to a solution at the same spectral value on the
background $u^{[r]}$.

Starting from the vacuum eigenfunction $\chi_j$ at
$\beta=\kappa_j^2$ and applying successively
$D_1,\ldots,D_{j-1}$, we define
\[
\phi_j:=D_{j-1}D_{j-2}\cdots D_1\chi_j,
\qquad j\geq2.
\]
The spatial intertwining relations
\[
L^{[r]}D_r=D_rL^{[r-1]}
\]
give
\[
\begin{aligned}
L^{[j-1]}\phi_j
&=
L^{[j-1]}D_{j-1}\cdots D_1\chi_j
\\
&=
D_{j-1}\cdots D_1L^{[0]}\chi_j
\\
&=
\kappa_j^2\phi_j.
\end{aligned}
\]
The temporal part follows successively from the scalar Darboux
property, yielding
\[
(\phi_j)_t
=
-\frac12(u^{[j-1]})_x\phi_j
+
\left(4\kappa_j^2+u^{[j-1]}\right)(\phi_j)_x.
\]
Hence $\phi_j$ satisfies the complete scalar Lax pair on the
background $u^{[j-1]}$ at the spectral value $\kappa_j^2$.

In particular,
\[
\phi_1=\chi_1,
\qquad
\phi_2=D_1\chi_2,
\qquad
\phi_3=D_2D_1\chi_3,
\]
and, more generally, the successive Darboux chain is
\[
u^{[0]}
\xrightarrow{\;\phi_1=\chi_1\;}
u^{[1]}
\xrightarrow{\;\phi_2=D_1\chi_2\;}
u^{[2]}
\xrightarrow{\;\phi_3=D_2D_1\chi_3\;}
\cdots
\xrightarrow{\;\phi_n=D_{n-1}\cdots D_1\chi_n\;}
u^{[n]} .
\]

The preceding bosonic construction extends naturally to the ghost sector.
Starting from the vacuum value $v^{[0]}=0$, each Darboux step transports
the previously constructed ghost field and introduces an additional
homogeneous contribution associated with the dressed eigenfunction
$\phi_j$. The resulting recursive construction is summarized below.

\begin{teorema}[Recursive ghost extension of the Darboux chain]
Let
\[
u^{[0]}\longrightarrow u^{[1]}\longrightarrow\cdots
\longrightarrow u^{[n]}
\]
be the bosonic Darboux chain constructed above, with
\[
p^{[j]}=(\log\phi_j)_x,
\qquad
u^{[j]}=u^{[j-1]}+4p^{[j]}_x,
\qquad
\alpha_j=\kappa_j^2.
\]
Fix $j\geq 1$ and work on a domain on which $\phi_j\neq 0$. Assume that
\(
\bigl(u^{[j-1]},v^{[j-1]}\bigr)
\)
satisfies the local KdV--ghost system, and let
$\eta^{[j]}_p$ be a particular solution of
\[
(\eta^{[j]}_p)_x
=
-\frac12 v^{[j-1]}
-2p^{[j]}\eta^{[j]}_p,\qquad
(\eta^{[j]}_p)_t
=
\left[
-\frac12(v^{[j-1]})_x
+p^{[j]}v^{[j-1]}
+\left(4\kappa_j^2+u^{[j-1]}\right)
\eta^{[j]}_p
\right]_x.
\]

Then every solution of the corresponding auxiliary ghost system is of
the form
\[
\eta^{[j]}
=
\eta^{[j]}_p+\zeta_j\phi_j^{-2},
\]
where $\zeta_j$ is an odd element of the underlying Grassmann algebra
independent of $x$ and $t$.

Defining
\[
v^{[j]}
=
v^{[j-1]}+4(\eta^{[j]})_x,
\]
the pair
\(
\bigl(u^{[j]},v^{[j]}\bigr)
\)
again satisfies the local KdV--ghost system.

Consequently, starting from $v^{[0]}=0$, the Darboux chain can be
lifted recursively to the ghost sector, yielding
\(
v^{[j]}
=
v^{[j]}(x,t;\zeta_1,\ldots,\zeta_j),
\quad
j=1,\ldots,n,
\)
with one new odd homogeneous parameter introduced at each Darboux step.
\end{teorema}

\begin{proof}
Fix $j\geq 1$ and work on a domain on which $\phi_j\neq 0$. By construction,
\[
p^{[j]}=(\log\phi_j)_x=\frac{(\phi_j)_x}{\phi_j},
\]
and $\phi_j$ satisfies the complete scalar Lax pair on the
background $u^{[j-1]}$ at the spectral value $\kappa_j^2$:
\[
(\phi_j)_{xx}
=
\left(
\kappa_j^2-\frac12u^{[j-1]}
\right)\phi_j,\qquad (\phi_j)_t
=
-\frac12(u^{[j-1]})_x\phi_j
+
\left(
4\kappa_j^2+u^{[j-1]}
\right)(\phi_j)_x.
\]
It follows that
\[
p^{[j]}_x
=
\kappa_j^2-\frac12u^{[j-1]}-(p^{[j]})^2,
\]
and, since $p^{[j]}=(\log\phi_j)_x$,
\[
p^{[j]}_t
=
\left[
-\frac12(u^{[j-1]})_x
+
\left(
4\kappa_j^2+u^{[j-1]}
\right)p^{[j]}
\right]_x.
\]
Thus $p^{[j]}$ satisfies the bosonic auxiliary system of
Theorem~3 with $\alpha=\kappa_j^2$.

Now assume that
\(
\bigl(u^{[j-1]},v^{[j-1]}\bigr)
\)
satisfies the local KdV--ghost system. By Lemma~3, the
corresponding auxiliary ghost system is compatible. Let
$\eta_p^{[j]}$ be a particular solution and let $\eta^{[j]}$
be any other solution. Setting
\(
\rho:=\eta^{[j]}-\eta_p^{[j]},
\)
the difference satisfies the homogeneous system
\[
\rho_x=-2p^{[j]}\rho,
\qquad
\rho_t=
\left[
\left(
4\kappa_j^2+u^{[j-1]}
\right)\rho
\right]_x.
\]

Using
\[
p^{[j]}=\frac{(\phi_j)_x}{\phi_j},
\]
the spatial equation gives
\[
(\phi_j^2\rho)_x
=
2\phi_j(\phi_j)_x\rho+\phi_j^2\rho_x
=
0.
\]
Let
\[
A_j:=4\kappa_j^2+u^{[j-1]}.
\]
From
\[
(\phi_j)_t
=
-\frac12(u^{[j-1]})_x\phi_j
+
A_j(\phi_j)_x,
\]
together with
\[
\rho_t=(A_j\rho)_x,
\qquad
(A_j)_x=(u^{[j-1]})_x,
\]
we also obtain
\[
\begin{aligned}
(\phi_j^2\rho)_t
&=
2\phi_j(\phi_j)_t\rho+\phi_j^2\rho_t
\\
&=
2\phi_j
\left[
-\frac12(u^{[j-1]})_x\phi_j
+
A_j(\phi_j)_x
\right]\rho
\\
&\quad
+
\phi_j^2
\left[
(u^{[j-1]})_x\rho+A_j\rho_x
\right]
\\
&=0,
\end{aligned}
\]
where in the last step we used
\[
\rho_x=-2p^{[j]}\rho
=
-2\frac{(\phi_j)_x}{\phi_j}\rho.
\]
Hence $\phi_j^2\rho$ is independent of both $x$ and $t$.

 Denoting this
odd constant by $\zeta_j$, we obtain
\[
\eta^{[j]}
=
\eta_p^{[j]}+\zeta_j\phi_j^{-2}.
\]

We now define
\[
v^{[j]}
=
v^{[j-1]}+4(\eta^{[j]})_x.
\]
The fields $p^{[j]}$ and $\eta^{[j]}$ satisfy the bosonic and
ghost auxiliary equations of Theorem~3, respectively, with
$\alpha=\kappa_j^2$. Therefore Theorem~3 gives
\[
u^{[j]}
=
u^{[j-1]}+4p^{[j]}_x,
\qquad
v^{[j]}
=
v^{[j-1]}+4(\eta^{[j]})_x,
\]
and guarantees that
\(
\bigl(u^{[j]},v^{[j]}\bigr)
\)
again satisfies the local KdV--ghost system.

For the initial step, $v^{[0]}=0$, and one may choose
\(
\eta_p^{[1]}=0.
\)
Hence
\[
\eta^{[1]}=\zeta_1\phi_1^{-2},
\qquad
v^{[1]}=4\zeta_1(\phi_1^{-2})_x.
\]
The preceding argument then provides the inductive step
\[
\bigl(u^{[j-1]},v^{[j-1]}\bigr)
\longmapsto
\bigl(u^{[j]},v^{[j]}\bigr),
\]
so the construction can be repeated for
$j=1,\ldots,n$.

Finally, the resulting family does not depend on the choice of
particular solution. Indeed, if $\widehat\eta_p^{[j]}$ is another
particular solution, then
\[
\widehat\eta_p^{[j]}-\eta_p^{[j]}
=
c\,\phi_j^{-2}
\]
for some $x$- and $t$-independent odd constant $c$. Hence
\[
\eta_p^{[j]}+\zeta_j\phi_j^{-2}
=
\widehat\eta_p^{[j]}
+
(\zeta_j-c)\phi_j^{-2},
\]
so changing the particular solution only redefines the homogeneous
parameter $\zeta_j$.
\end{proof}

Consequently, for arbitrary $x$- and $t$-independent odd constants
$\vartheta_1$ and $\vartheta_2$,
\[
v^{[2]}
=
\vartheta_1\frac{\partial u^{[2]}}{\partial\delta_1}
+
\vartheta_2\frac{\partial u^{[2]}}{\partial\delta_2}
\]
is a local ghost field on the same two-step background. Together with
the one-soliton result of Subsection~6.1, this shows explicitly that
variations of the soliton phase parameters provide natural tangent
directions in the local ghost sector. This interpretation is consistent
with the recursive Darboux construction of Theorem~4, in which a new
odd homogeneous freedom appears at each successive step.

\section{Discussion and conclusions}

The preceding results place the KdV--ghost system at the intersection
of several structures that arise naturally in the theory of integrable
systems.

We analyzed the existence of a shift symmetry on the ghost field of the
KdV--ghost system. Its generator does not commute with the generator of the
BRST symmetry; the commutator is a spatial translation on the fields with
parameter given by the product of the nilpotent parameters of the BRST and
shift transformations.

By constructing a sequence of
BRST $\rightarrow$ shift $\rightarrow$ BRST transformations,
we showed how to obtain new solutions starting from a known one.

We obtained a Lax pair and a zero-curvature formulation for the KdV--ghost
system and analyzed the spectral problem of its spatial Lax operator, showing
an isospectral theorem and the relation to the spectral problem of the spatial
Lax operator of the KdV equation.

The domain of the operators is defined on a Grassmann algebra with coefficients
in the Sobolev space $H^2$. It turns out that the spatial Lax operator of the
KdV--ghost system is a ``nilpotent'' conjugation of the corresponding Lax
operator of the KdV equation, an interesting property which we strongly used
in the analysis.

In the local variable $v=C_x$, the ghost equation coincides
with the Fr\'echet linearization of the KdV flow, whereas its origin
remains intrinsically tied to the Grassmann-valued BRST formulation.
At the same time, the nilpotent gauge representation and the
Darboux construction show that this tangent interpretation is compatible
with the spectral and solution-generating structures of the bosonic
KdV equation. We discuss below the relation of these results with
integrable couplings and supersymmetric extensions of KdV, and then
summarize the geometric interpretation that emerges from the
construction.

\subsection{Relation with integrable couplings and supersymmetric KdV}

The local KdV--ghost system
\[
u_t=K[u],\qquad v_t=K'[u]v,
\]
has the same differential form as the first-order perturbation systems
associated with the KdV hierarchy. Such systems were studied by Ma and
Fuchssteiner in connection with the bi-Hamiltonian structure of
perturbation equations and, more generally, by Ma in the framework of
integrable couplings \cite{MaFuchssteiner1996,Ma2000}. From this point
of view, the field $v$ describes the tangent dynamics along a KdV
solution $u$.

The same differential system also appears, up to a simple
normalization, in Sakovich's classification of coupled KdV equations
of Hirota--Satsuma type \cite{Sakovich1999}. Indeed, under the change
of variables
\[
U=-\frac14 u,\qquad V=v,
\]
the local KdV--ghost equations become
\[
U_t=U_{xxx}-12UU_x,\qquad
V_t=V_{xxx}-12VU_x-12UV_x,
\]
which is precisely case (ix) of Sakovich for $k=0$. Sakovich also
constructs a Lax representation for this coupled system. The coincidence concerns the differential form of the equations.

In the present construction, however, the field $v$ has a different
origin. It is Grassmann odd and arises as the local variable $v=C_x$
associated with the BRST ghost field $C$. Thus, although the local
equations have the differential form of a perturbative integrable
coupling, the odd sector retains the BRST structure of the original
KdV--ghost system, including the ghost-shift symmetry and the
nilpotent spectral construction developed above.

The presence of a Grassmann-odd field also invites comparison with
supersymmetric extensions of the KdV hierarchy \cite{ManinRadul1985,Mathieu1988}.
The present system, however, should not be identified with
supersymmetric KdV: the bosonic field satisfies the ordinary KdV
equation, while the odd field obeys its Fr\'echet linearization and
originates from the BRST ghost sector rather than from a spacetime
supersymmetry multiplet. The closure of the BRST and ghost-shift
transformations on a spatial translation should therefore be understood
as an algebraic resemblance to supersymmetric closure relations, rather
than as a supersymmetric extension of KdV.

\subsection{Geometric interpretation of the ghost sector}

The passage from the ghost representative $C$ to the local variable
$v=C_x$ gives a direct geometric interpretation of the odd sector. At
the representative level, fields that differ by an $x$-independent
odd function determine the same local field $v$. After passing to this local variable, the ghost equation becomes
precisely the Fr\'echet linearization of the KdV flow,
\[
v_t=K'[u]v,
\]
so that $v$ defines a tangent direction along the KdV trajectory
determined by $u$. This tangent interpretation is local and does not
replace the BRST origin of the ghost field; rather, it describes the
differential content that remains after passing from the representative
$C$ to $v=C_x$.

For the solitonic solutions, this interpretation is realized
explicitly by variations of the phase parameters. In the one-soliton
case the homogeneous odd freedom is proportional to the phase
derivative of the KdV solution, while for the two-step Darboux solution
the corresponding phase variations again produce local ghost fields.
The recursive Darboux construction of Theorem~4 extends this picture:
at each step a new odd homogeneous freedom appears, associated with
the dressed eigenfunction generating that Darboux transformation.
Thus the ghost sector may be viewed simultaneously as a BRST-induced
odd extension and, in the local formulation, as a tangent extension of
the bosonic KdV dynamics.

Taken together, these results provide a local Darboux description of
the KdV--ghost sector that is compatible with its BRST origin, its
spectral formulation, and its tangent interpretation.


\section*{Conflict of interest}

The authors state that there is no conflict of interest.

\begin{appendices}
\numberwithin{equation}{section}

\section{The Lax pair for the KdV-ghost system}

In this appendix we derive the Grassmann-valued differential Lax
operators introduced in Subsection~3.1. We begin with the
differential-operator ansatz
\begin{equation}
\mathscr{L}
=
D_x^2+F_0D_x+F_1I,
\qquad
\mathscr{M}
=
g_0D_x^3+g_1D_x^2+g_2D_x+g_3I,
\label{eq:app-lax-ansatz}
\end{equation}
where $g_0\in\mathbb{R}$ is constant and
$F_0,F_1,g_1,g_2,g_3$ are even Grassmann-valued coefficient functions.

We impose the Lax equation
\begin{equation}
\partial_t\mathscr{L}
=
[\mathscr{M},\mathscr{L}],
\label{eq:app-lax-equation}
\end{equation}
and compare the coefficients of the different powers of $D_x$.

Substituting $\mathscr{L}$ and $\mathscr{M}$ into
\eqref{eq:app-lax-equation}, the coefficient of $D_x^3$ gives
\begin{equation*}
3g_0(F_0)_x-2(g_1)_x=0.
\label{eq:app-D3}
\end{equation*}
Hence,
\[
g_1=\frac{3}{2}g_0F_0,
\]
where the $x$-independent integration term has been set to zero.

The coefficient of $D_x^2$ gives
\begin{equation}
3g_0(F_0)_{xx}
+3g_0(F_1)_x
-(g_1)_{xx}
+2g_1(F_0)_x
-F_0(g_1)_x
-2(g_2)_x
=0.
\label{eq:app-D2}
\end{equation}

We now take
\begin{equation}
F_0=\epsilon C_x,
\qquad
\epsilon^2=0,
\label{eq:app-F0}
\end{equation}
where $\epsilon$ and $C$ are odd Grassmann-valued quantities. Using the expression for $g_1$ obtained above, we obtain
\begin{equation}
g_2
=
\frac{3}{4}g_0(F_0)_x
+
\frac{3}{2}g_0F_1,
\label{eq:app-g2}
\end{equation}
where again the $x$-independent integration term has been set to zero.

The coefficient of $D_x$ becomes
\begin{equation}
(F_0)_t
=
\frac{g_0}{4}(F_0)_{xxx}
+
\frac{3}{2}g_0(F_1)_{xx}
+
\frac{3}{2}g_0D_x(F_0F_1)
-
2(g_3)_x.
\label{eq:app-D1}
\end{equation}

We choose
\begin{equation*}
g_3
=
\frac{3}{4}g_0(F_1)_x,
\label{eq:app-g3}
\end{equation*}
so that
\begin{equation}
(F_0)_t
=
\frac{g_0}{4}(F_0)_{xxx}
+
\frac{3}{2}g_0D_x(F_0F_1).
\label{eq:app-F0-evolution}
\end{equation}

Finally, we fix
\begin{equation*}
g_0=4,
\qquad
F_1
=
\frac{1}{2}
\left(
u+\epsilon C_{xx}
\right).
\label{eq:app-F1}
\end{equation*}
Since
\[
F_0F_1
=
\frac{\epsilon}{2}uC_x,
\]
equation \eqref{eq:app-F0-evolution} yields
\[
\epsilon D_x
\left(
C_t-C_{xxx}-3uC_x
\right)
=
0.
\]
Therefore,
\begin{equation}
D_x
\left(
C_t-C_{xxx}-3uC_x
\right)
=
0.
\label{eq:app-ghost}
\end{equation}
Equivalently,
\begin{equation*}
C_t
=
C_{xxx}
+
3uC_x
+
\Gamma(t),
\label{eq:app-gamma}
\end{equation*}
where $\Gamma(t)$ is odd and independent of $x$.

Finally, the coefficient of the identity operator gives
\begin{equation}
(F_1)_t
=
g_0(F_1)_{xxx}
+
g_1(F_1)_{xx}
+
g_2(F_1)_x
-
(g_3)_{xx}
-
F_0(g_3)_x.
\label{eq:app-I}
\end{equation}

Substituting the expressions obtained above and separating the terms
independent of $\epsilon$ from those linear in $\epsilon$ gives,
respectively,
\begin{equation}
u_t-u_{xxx}-3uu_x=0,
\label{eq:app-kdv}
\end{equation}
and
\begin{equation*}
D_x^2
\left(
C_t-C_{xxx}-3uC_x
\right)
=
0.
\label{eq:app-ghost-second}
\end{equation*}
The latter is already implied by \eqref{eq:app-ghost}.

Therefore, the resulting Grassmann-valued differential Lax operators
are
\begin{equation}
\mathscr{L}_\epsilon
=
D_x^2
+
\epsilon C_xD_x
+
\frac{1}{2}
\left(
u+\epsilon C_{xx}
\right)I,
\label{eq:app-Lepsilon}
\end{equation}
and
\begin{equation}
\mathscr{M}_\epsilon
=
4D_x^3
+
6\epsilon C_xD_x^2
+
\left(
3u+6\epsilon C_{xx}
\right)D_x
+
\frac{3}{2}
\left(
u_x+\epsilon C_{xxx}
\right)I.
\label{eq:app-Mepsilon}
\end{equation}
These are precisely the Grassmann-valued Lax operators introduced in
Subsection~3.1.

\section{Solitonic and Hirota-type solutions}
\label{app:previous-solutions}

For completeness, we collect here several representative-level ghost
solutions obtained before introducing the local variable $v=C_x$ and
the Darboux formulation developed in the main text. These results are
not required for the proof of the nilpotent Darboux transformation,
but they clarify how the earlier recursive, travelling-wave, and
Hirota expressions fit into the local framework.

Throughout this appendix,
\(
\zeta\in\Lambda_1
\)
denotes an $x$- and $t$-independent odd element of the Grassmann
algebra.

\subsection{The recursive one-soliton representative}

A representative obtained from the recursive construction is
\begin{equation}
C
=
\left(
u_{xx}+\frac{3}{2}u^2
\right)\zeta .
\label{eq:app-recursive-ghost}
\end{equation}
A direct substitution of
\[
u_t=u_{xxx}+3uu_x
\]
shows that \eqref{eq:app-recursive-ghost} satisfies the representative
ghost equation
\[
C_t=C_{xxx}+3uC_x
\]
whenever $u$ satisfies the KdV equation.

Consider now the one-soliton solution
\begin{equation*}
u(x,t)
=
c\,\operatorname{sech}^2\xi,
\qquad
\xi
=
\frac{\sqrt{c}}{2}
\left(
x+ct+x_0
\right),
\qquad
c>0.
\label{eq:app-one-soliton}
\end{equation*}
A direct calculation gives
\begin{equation*}
u_{xx}
=
cu-\frac{3}{2}u^2.
\label{eq:app-one-soliton-identity}
\end{equation*}
Hence
\(
C=cu\,\zeta.
\label{eq:app-one-soliton-ghost}
\) Since the representative ghost equation is linear and homogeneous in
$C$, more generally
\(
C
=
\mu u\,\zeta,
\quad
\mu\in\mathbb{R},
\label{eq:app-one-soliton-family}
\)
is a solution for every real constant $\mu$. The corresponding local
ghost field is
\(
v
=
C_x
=
\mu u_x\,\zeta.
\label{eq:app-one-soliton-local}
\)
Thus the localized representative family reduces, after passing to
$v=C_x$, to the translation tangent mode of the one-soliton solution,
in agreement with the vacuum-to-soliton Darboux construction of
Subsection~6.1.

\subsection{Travelling-wave ghost representatives}

We retain the one-soliton background introduced in Subsection B.1 and seek representative ghost solutions of the form
\(
C(x,t)
=
F(\xi)\zeta.
\label{eq:app-travelling-ansatz}
\)
Substitution into
\[
C_t=C_{xxx}+3uC_x
\]
gives
\begin{equation}
F_{\xi\xi\xi}
+
\left(
12\operatorname{sech}^2\xi-4
\right)F_\xi
=
0.
\label{eq:app-travelling-third}
\end{equation}
Defining
\(
w:=F_\xi,
\)
we obtain
\begin{equation*}
w_{\xi\xi}
+
\left(
12\operatorname{sech}^2\xi-4
\right)w
=
0.
\label{eq:app-poschl-teller}
\end{equation*}
This is a P\"oschl--Teller-type equation with parameter $\nu=3$.

A convenient form of the resulting family is
\begin{equation}
F(\xi)
=
a_0
-\frac{a_1}{2}\operatorname{sech}^2\xi
+a_2H(\xi),
\qquad
a_0,a_1,a_2\in\mathbb{R},
\label{eq:app-travelling-family}
\end{equation}
where
\begin{equation*}
\begin{split}
H(\xi)
={}&
-\frac{4}{5}\xi
-\frac{1}{30}e^{2\xi}
+\frac{1}{30}e^{-2\xi}
-2\,
\frac{-2\xi e^{2\xi}+e^{2\xi}+1}
     {(e^{2\xi}+1)^2}.
\end{split}
\label{eq:app-unbounded-branch}
\end{equation*}
A direct substitution verifies that
\eqref{eq:app-travelling-family} satisfies
\eqref{eq:app-travelling-third}.

The constant term $a_0$ represents the freedom to add an
$x$-independent odd element to the representative $C$ and therefore
disappears after passing to the local field $v=C_x$. The term
proportional to $a_1$ is localized and belongs to the family already
obtained in \eqref{eq:app-one-soliton-family}, since
\[
-\frac{a_1}{2}\operatorname{sech}^2\xi
=
-\frac{a_1}{2c}\,u(\xi).
\]

By contrast,
\begin{equation*}
H(\xi)
\sim
-\frac{1}{30}e^{2\xi},
\qquad
\xi\to+\infty,
\end{equation*}
and
\begin{equation*}
H(\xi)
\sim
\frac{1}{30}e^{-2\xi},
\qquad
\xi\to-\infty.
\end{equation*}
Thus $H$ is unbounded at both spatial ends. Its derivative is also
unbounded, so for $a_2\neq0$ neither the representative $C$ nor the
corresponding local ghost field $v=C_x$ belongs to the Schwartz class.
This branch therefore defines a formal travelling-wave solution in a
broader functional class, rather than a localized Darboux-generated
profile.

\subsection{Hirota representation and the temporal tangent mode}

Let $\tau=\tau(x,t)$ be a KdV tau function in the normalization used
here, following Hirota's bilinear formalism~\cite{Hirota2004}, so that
\begin{equation}
u=4\partial_x^2\log\tau.
\end{equation}
To distinguish the Hirota bilinear derivative from the total
derivative $D_x=\partial_x$, we denote the former by
$\mathcal{D}_x$:
\begin{equation}
\mathcal{D}_x^m(f\cdot g)
:=
\left.
(\partial_x-\partial_{x'})^m
f(x)g(x')
\right|_{x'=x}.
\label{eq:app-hirota-derivative}
\end{equation}

Writing $g=\log\tau$, one has
\begin{equation}
\frac{\mathcal{D}_x^4(\tau\cdot\tau)}{\tau^2}
=
2\left(
g_{xxxx}+6g_{xx}^2
\right).
\label{eq:app-hirota-identity}
\end{equation}
Using $u=4g_{xx}$, it follows that
\begin{equation*}
u_{xx}
+\frac{3}{2}u^2
=
2\,
\frac{\mathcal{D}_x^4(\tau\cdot\tau)}{\tau^2}.
\label{eq:app-hirota-kdv}
\end{equation*}
Consequently, the representative
\eqref{eq:app-recursive-ghost} can be written as
\begin{equation}
C
=
2\,
\frac{\mathcal{D}_x^4(\tau\cdot\tau)}{\tau^2}
\,\zeta.
\label{eq:app-hirota-ghost}
\end{equation}

For example, the two-soliton tau function may be written as
\begin{equation}
\tau
=
1+e^{\vartheta_1}
+e^{\vartheta_2}
+A_{12}e^{\vartheta_1+\vartheta_2},
\label{eq:app-two-soliton-tau}
\end{equation}
where
\begin{equation}
\vartheta_i
=
k_i x+k_i^3t+\vartheta_{i,0},
\qquad
A_{12}
=
\left(
\frac{k_1-k_2}{k_1+k_2}
\right)^2.
\label{eq:app-two-soliton-data}
\end{equation}
Equations~(B.6) and
\eqref{eq:app-hirota-ghost} therefore give a representative
KdV--ghost solution on the two-soliton background.

The relation with the local tangent formulation is immediate. The KdV
equation can be written in conservation form as
\begin{equation*}
u_t
=
D_x
\left(
u_{xx}+\frac{3}{2}u^2
\right).
\label{eq:app-kdv-conservation}
\end{equation*}
Hence
\(
C_x
=
\zeta u_t.
\label{eq:app-hirota-local}
\)
Thus the corresponding local ghost field is the odd temporal tangent
direction of the KdV solution.

For the two-soliton parametrization used in Subsection~6.2,
\(
\xi_i
=
\kappa_i x+4\kappa_i^3t+\delta_i,
\)
one has
\[
u_t
=
4\kappa_1^3
\frac{\partial u}{\partial\delta_1}
+
4\kappa_2^3
\frac{\partial u}{\partial\delta_2}.
\]
Therefore
\(
v
=
C_x
=
\zeta u_t
\)
belongs to the phase-variation family obtained from the successive
Darboux construction. In this sense, the earlier Hirota
representative is consistent with the tangent and Darboux
interpretations developed in the main text.

\end{appendices}

\end{document}